\documentclass[letterpaper,twocolumn,10pt]{article}
\usepackage{usenix}
\usepackage{amsmath}
\usepackage{amsfonts}
\usepackage{enumitem}
\usepackage{algorithm}
\usepackage{algpseudocode}
\usepackage{tabularx}
\usepackage{booktabs}
\usepackage{caption}
\usepackage{xurl}
\usepackage{graphicx}
\usepackage{textcomp}
\usepackage{multirow} 
\usepackage{xspace}
\usepackage{filecontents}
\usepackage[english]{babel}
\usepackage{blindtext}
\usepackage{tikz}
\definecolor{comment}{rgb}{0.54,0.1,0.066}
\usepackage{todonotes}
\usepackage[final]{changes}
\definechangesauthor[name={AU}, color=blue]{AU} 
\newcommand{\blackcirclednum}[1]{%
  \tikz[baseline=(char.base)]{%
    \node[shape=circle, fill=black, text=white, inner sep=0.5pt, font=\footnotesize] (char) {#1};%
  }%
}

\begin{document}

\date{}

\newcommand{\sys}{\textsf{FENIX}\xspace}
\newcommand{\argmax}{\textsf{argmax}\xspace}

\title{FENIX: Enabling In-Network DNN Inference with \\ FPGA-Enhanced Programmable Switches}

\author{
\begin{tabular}{cccc}
Xiangyu Gao\textsuperscript{\S} & 
Tong Li\textsuperscript{\ddag} & 
Yinchao Zhang\textsuperscript{\S} & 
Ziqiang Wang\textsuperscript{\dag} \\
\multicolumn{4}{c}{
Xiangsheng Zeng\textsuperscript{$\star$} \quad
Su Yao\textsuperscript{\S, \color{green!80!black}{$\ast$}} \quad
Ke Xu\textsuperscript{\S,}\thanks{Ke Xu and Su Yao are the corresponding authors.}
}
\end{tabular}
\\[1.5ex]
\textsuperscript{\S}Tsinghua University \quad
\textsuperscript{\ddag}Renmin University of China \\
\textsuperscript{\dag}Southeast University \quad
\textsuperscript{$\star$}Huazhong University of Science and Technology
}
\maketitle

\begin{abstract}
Machine learning (ML) is increasingly used in network data planes for advanced traffic analysis, but existing solutions (such as FlowLens, N3IC, BoS) still struggle to simultaneously achieve low latency, high throughput, and high accuracy. To address these challenges, we present \sys, a hybrid in-network ML system that performs feature extraction on programmable switch ASICs and deep neural network inference on FPGAs. \sys introduces a Data Engine that leverages a probabilistic token bucket algorithm to control the sending rate of feature streams, effectively addressing the throughput gap between programmable switch ASICs and FPGAs. In addition, \sys designs a Model Engine to enable high-accuracy deep neural network inference in the network, overcoming the difficulty of deploying complex models on resource-constrained switch chips. We implement \sys on a programmable switch platform that integrates a Tofino ASIC and a ZU19EG FPGA directly, and evaluate it on real-world network traffic datasets. Our results show that \sys achieves microsecond-level inference latency and multi-terabit throughput with low hardware overhead, and delivers over 90\% accuracy on mainstream network traffic classification tasks, outperforming the state of the art.

\end{abstract}

\vspace{-1em}
\section{Introduction}
\vspace{-1em}
Machine learning (ML) is increasingly transforming networked systems. Recent work demonstrates that both traditional ML techniques—such as decision trees, random forests, and support vector machines—and deep learning (DL) approaches can significantly improve networking tasks, including malicious traffic detection~\cite{kitsune, fu2021realtime, DBLP:conf/uss/netbeacon, jafri_leo_2024} and application traffic classification~\cite{DBLP:conf/ndss/flowlens, DBLP:conf/uss/netbeacon, yan2024brain}. These advances have been enabled by the emergence of programmable network devices, including P4 switches~\cite{bosshart_p4_2014}, NetFPGA~\cite{noauthor_netfpga_nodate}, and SmartNICs~\cite{firestone2018azure}. By integrating data-driven learning models with traditional rule-based mechanisms, these systems enable more adaptive and responsive in-network traffic analysis.

Early intelligent network designs typically split responsibilities between planes: the programmable data plane extracts traffic features, while the control plane performs ML inference. Although this leverages existing programmable infrastructure, it introduces communication latencies—often milliseconds~\cite{DBLP:conf/ndss/flowlens,swamy_taurus_2022}. Such delays are especially problematic for time-sensitive applications like intrusion detection, as analysis may lag behind real-time traffic (see \S~\ref{sec:evaluation} for details).

To address these latency challenges, researchers have explored moving ML inference closer to the data path. One promising approach is implementing inference directly on programmable devices such as SmartNICs, as demonstrated by systems like N3IC~\cite{siracusano_re-architecting_2022}. This offloads computation from endpoint CPUs and significantly reduces inference latency. However, new bottlenecks arise: for example, N3IC achieves maximum throughput of only 40 Gbps, and even the latest commercial SmartNICs are limited to 400 Gbps~\cite{nvidia_connectx7_2025}. Such throughput is often insufficient for core networks, where switch ASICs handle 4 to 12Tbps\cite{tofino2, tofino}, far exceeding current SmartNIC capabilities. While deploying multiple NICs can scale throughput, this approach introduces significant costs and new scheduling challenges, as packet scheduling consumes significant CPU resources~\cite{gupta_loom_2019}.

Recent advances in in-network ML have led researchers to implement ML models directly on programmable switch ASICs. For example, systems such as Leo~\cite{jafri_leo_2024} and NetBeacon~\cite{DBLP:conf/uss/netbeacon} deploy decision tree and random forest models on Tofino switches using Match Action Tables (MAT), while BoS~\cite{yan2024brain} demonstrates the feasibility of running binary recurrent neural networks (RNNs) on similar hardware. These approaches can potentially enable line-rate ML inference at significantly higher throughput. However, they are typically subject to stringent computational resource constraints, which limit their effectiveness for complex multi-classification tasks (see \S~\ref{sec:evaluation} for details).

These observations raise several key questions: Why do existing switch ASICs struggle to support more complex ML tasks? Beyond the programmable devices discussed above, what other platforms can balance accuracy and latency for in-network ML? In this paper, we present \sys (FPGA-Enabled Neural Inference eXecution for network switches), a system designed to achieve low latency, high throughput, and high accuracy for in-network intelligent traffic analysis.

\begin{figure}[t]
\centering
\includegraphics[width=0.48\textwidth]{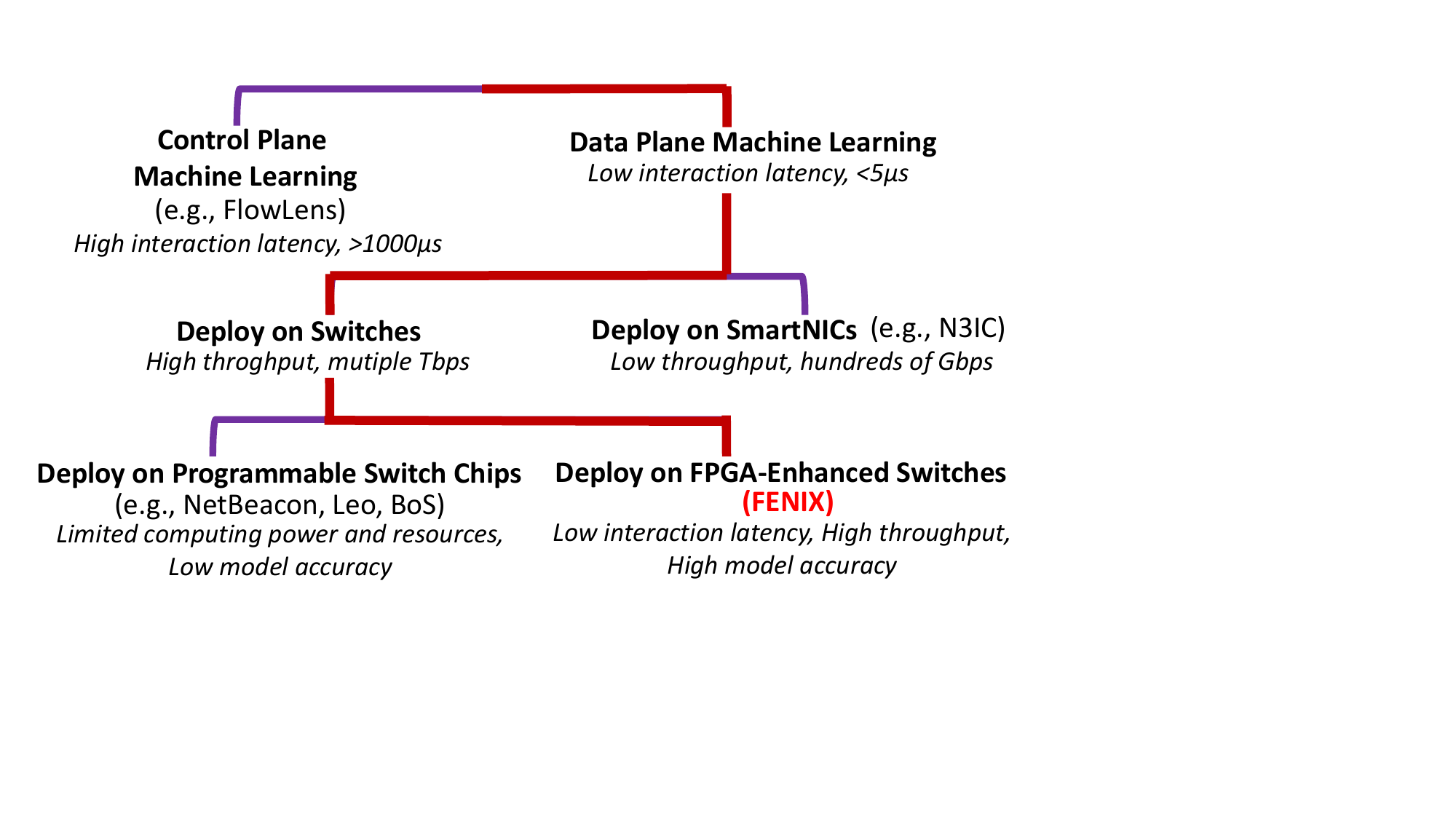}
\caption{Design space in Intelligent Network.}
\label{fig:designtree}
        \vspace{-0.7cm}
\end{figure}

Figure~\ref{fig:designtree} surveys intelligent network deployment methods and positions \sys within this landscape. Unlike Control Plane approaches such as FlowLens~\cite{DBLP:conf/ndss/flowlens}, which incur high interaction latency (>1000$\mu$s), \sys adopts data plane Machine Learning to achieve sub-microsecond latency (<5$\mu$s). Within the data plane category, \sys is implemented on switches rather than SmartNICs (e.g., N3IC~\cite{siracusano_re-architecting_2022}), supporting multi-terabit-per-second throughput compared to only hundreds of Gbps with SmartNIC-based solutions. Furthermore, \sys leverages FPGA-enhanced switches instead of solely relying on programmable switch ASICs (e.g., NetBeacon~\cite{DBLP:conf/uss/netbeacon}, Leo~\cite{jafri_leo_2024}, BoS~\cite{yan2024brain}), thereby overcoming their computational and resource limitations and enabling higher model accuracy while retaining low latency and high throughput.

\sys consists of two main components: the Data Engine and Model Engine, which enable in-network traffic analysis on the data plane. The Data Engine extracts features from high-throughput traffic, while the Model Engine performs ML inference on these features. This design enables \sys to achieve three key goals: \textit{low latency} by avoiding software processing, \textit{high throughput} through efficient feature extraction, and \textit{high accuracy} by supporting full DNN models with minimal quantization loss.

\noindent\textbf{Contributions.} The main contribution of this paper is the design, implementation, and evaluation of \sys, the first system to leverage FPGA augmentation to extend the capabilities of programmable switches for in-network machine learning. Over six months, we designed and manufactured a custom switch, integrating a programmable Tofino switch chip and a ZU19EG FPGA on Printed Circuit Board (PCB). We evaluate \sys on challenging tasks, including VPN encrypted traffic classification and malware detection. Experimental results show that \sys achieves up to 537$\times$ lower inference latency compared to control plane-based approaches, and improves classification accuracy by up to 21\% over SOTA. Our prototype shows low hardware resource overhead, indicating that FPGA-enhanced switches are practical for deployment in high-speed network environments.

\vspace{-1.5em}
\section{BACKGROUND AND MOTIVATION}
\label{sec:background}
\vspace{-0.5em}
This section reviews the evolution of DNN models for networking, examines the limitations of programmable switches, highlights the promise of FPGA-assisted networking, and outlines the integration challenges that motivate our hybrid system design.

\noindent\textbf{DNN Learning Models for Networking.}
Classification tasks such as malicious traffic detection~\cite{kitsune, fu2021realtime,DBLP:conf/uss/netbeacon, jafri_leo_2024} and application traffic classification~\cite{DBLP:conf/ndss/flowlens, DBLP:conf/uss/netbeacon, yan2024brain} are fundamental for network intelligence. Prior work has demonstrated the feasibility of in-network AI inference by deploying lightweight models—such as decision trees~\cite{jafri_leo_2024}, random forests~\cite{DBLP:conf/uss/netbeacon}, XGBoost~\cite{DBLP:conf/uss/netbeacon}, binarized MLPs~\cite{siracusano_re-architecting_2022}, and simplified RNNs~\cite{yan2024brain}—on programmable switches and SmartNICs. While these models enable practical deployment, their limited representational power often makes it difficult to capture complex or dynamic network behaviors. In contrast, more complex DNNs offer stronger nonlinear modeling capabilities and can adapt more effectively to evolving network conditions, leading to improved accuracy on diverse networking tasks. As a result, supporting DNN inference in network devices has the potential to further enhance the flexibility and precision of in-network intelligence, especially in scenarios where lightweight models face limitations.

\noindent\textbf{Programmable Network Data Plane Limitations.}
Despite the promise of in-network DNN inference, the resource constraints of programmable data planes present significant challenges for deploying sophisticated models. PISA-based architectures are effective for basic machine learning tasks~\cite{DBLP:conf/uss/netbeacon, jafri_leo_2024}, but encounter fundamental limitations with complex neural networks. The PISA instruction set supports only simple operations (e.g., addition, subtraction, bit-shifting, and logical operations), and lacks native support for floating-point arithmetic, multiplication, division, and complex conditionals—operations essential for modern DNNs. Hardware resources are also limited: for example, commercial switches such as Barefoot Tofino 1 offer only 12 pipeline stages and constrained memory resources (120 Mbit SRAM, 6.2 Mbit TCAM)~\cite{yan2024brain}, which are insufficient for parameter-rich neural networks. Furthermore, restrictions such as atomic register access prevent efficient implementation of iterative computations required by DNNs. Collectively, these constraints make it impractical to support DNN inference on programmable switches while maintaining line-rate performance, underscoring the need for alternative approaches to bring advanced neural network capabilities into the data plane.

\noindent\textbf{FPGA-assisted Networking.}
FPGAs offer a promising solution to the computational limitations of programmable data planes while preserving high-performance networking capabilities. These reconfigurable platforms can interface directly with programmable switch ASICs, providing microsecond-level latency for DNN-based processing via customized hardware acceleration \footnote{An FPGA can be directly connected to a switching ASIC as an external module, as implemented in our system; this approach also applies to switching ASICs other than Tofino.}. FPGAs strike a favorable balance among energy efficiency, computational flexibility, and performance—surpassing CPUs in energy efficiency and providing more versatile analytical capabilities than fixed-function switch ASICs~\cite{zeng2022tiara}. Their support for partial dynamic reconfiguration enables network operators to update analytical functions without disrupting service~\cite{feng2024f3}, offering valuable operational flexibility. In hybrid architectures with programmable switch ASICs, these advantages are particularly valuable: conventional packet processing remains in the high-speed switch pipeline, while computation-intensive neural network inference is offloaded to specialized FPGA hardware. This clear division of labor enables the practical deployment of DNN models even in resource-constrained network environments.

\noindent\textbf{Mismatch between FPGA and Switch.}
Despite their complementary capabilities, fundamental architectural differences between programmable switches and FPGAs present significant integration challenges. The two components operate in fundamentally different ways: switches process packets in deterministic pipelines with nanosecond precision, whereas FPGAs perform variable-latency neural computations requiring sophisticated synchronization. More critically, there is a substantial throughput gap between these chips: modern programmable switches operate at multi-terabit-per-second rates~\cite{tofino}, while even the latest commercial FPGAs sustain only hundreds of Gbps~\cite{amd_zynq_ultrascale_plus_mpsoc}. This order-of-magnitude disparity makes it infeasible to offload all switch traffic directly to FPGAs for processing. As a result, a central challenge in designing FPGA-assisted networking systems is to efficiently extract and route only the most relevant feature information from high-speed network streams to the FPGA for deep processing, while preserving overall throughput and low latency.

\noindent\textbf{Motivation.}
Current research in-network traffic analysis mainly follows two directions: adapting machine learning models to fit data plane constraints (e.g., transforming decision trees into match-action table representations)~\cite{yan2024brain, DBLP:conf/uss/netbeacon, jafri_leo_2024, xie_mousika_2022, lee_switchtree_2020}, or sampling data for offline analysis on servers~\cite{DBLP:conf/ndss/flowlens, zheng_iisy_2022, kim2025sketchfeature}. However, both approaches inevitably compromise either model accuracy or introduce significant latency, and thus fall short of modern network requirements. The core challenge is to deploy powerful deep neural networks (DNNs) in high-speed network environments. While DNNs excel at traffic classification and anomaly detection, they face a critical bottleneck: programmable switches offer limited computational resources and restricted instruction sets, making complex DNN execution infeasible; meanwhile, hardware accelerators such as FPGAs, though efficient for neural network computation, provide processing bandwidth orders of magnitude lower than the Tbps-level throughput of switches. Even the most advanced FPGAs cannot directly process network traffic from switching chips. Therefore, it is necessary to select key feature information for the FPGA and implement rate control to adapt to different traffic rates.

\noindent\textbf{Goals.}
Our research aims to develop \sys, an intelligent network data plane system that simultaneously achieves three core technical goals:

• \noindent\textbf{Low latency:} \sys uses FPGA parallelism and direct interfacing with programmable switching ASICs to support neural network inference with microsecond-level latency, without involving software stacks or system buses. This hardware-based approach minimizes processing overhead and maximizes data throughput. As a result, \sys ensures real-time responsiveness for latency-sensitive network applications.

• \noindent\textbf{High throughput:} \sys achieves high throughput in two aspects. First, programmable switching ASICs process all network traffic at line rate, performing initial feature extraction without affecting forwarding performance. Second, a novel feature control mechanism optimizes the communication channel between switches and FPGAs by dynamically adjusting per-flow sampling rates. This approach maximizes effective bandwidth utilization while respecting hardware constraints.

• \noindent\textbf{High accuracy:} \sys implements different DNNs (e.g. CNN, RNN) on FPGAs using fixed-point quantization, rather than simplified or binarized models. This approach maintains accuracy close to offline analysis and enables support for advanced pattern recognition and temporal analysis tasks.
\begin{figure}[t]
    \centering
    \includegraphics[width=0.48\textwidth]{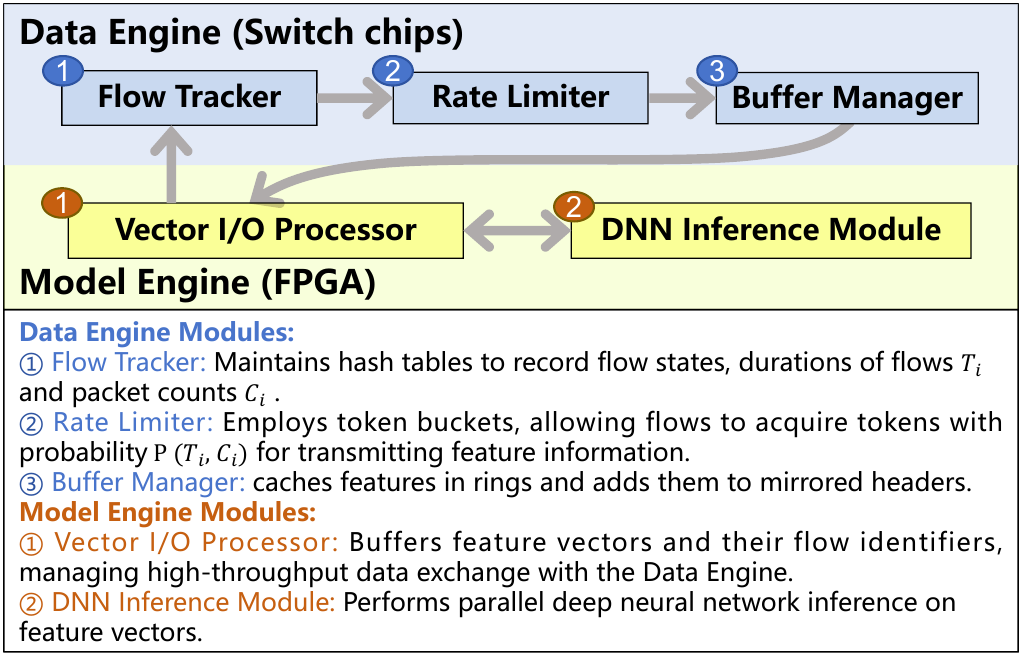}
    \caption{The architecture of \sys.}
    \label{fig:arch}
    \vspace{-0.7cm}
\end{figure}

\vspace{-1.2em}
\section{\sys Design Overview}
\label{sec:design}

Figure~\ref{fig:arch} shows the architecture of \sys, which is designed to address two main challenges in integrating deep learning inference with programmable switches. The first challenge is enabling the data plane to execute deep neural network models for fine-grained traffic control. The second challenge is bridging the throughput gap between FPGAs and programmable switching ASICs.

To address the throughput mismatch, \sys designs a Data Engine that manages feature caching and transmission control. The Flow Tracker is motivated by the need to enable per-flow precise inference, providing the foundation for applying DNN models to individual network flows. The Rate Limiter employs a token bucket mechanism with probabilistic control to regulate the rate at which each flow sends feature information, preventing the FPGA from being overwhelmed by excessive feature data. The Buffer Manager uses a ring buffer structure with mirrored headers, enabling the data engine to pass stateful feature information to the model engine for processing.

To support DNN inference on the data path, \sys designs a Model Engine responsible for both flow identification and model execution. The Vector I/O Processor is designed to support high-throughput, low-latency data exchange between the data engine and model engine, and caches flow identifiers so that the ASIC can accurately identify inference results. The DNN Inference Module performs parallel deep neural network inference on feature vectors, maximizing FPGA resource utilization.

With this design, \sys achieves high-throughput, low-latency, and accurate DNN inference for in-network traffic analysis. The following sections detail each module design.

\begin{figure}[t]
    \centering
    \includegraphics[width=0.48\textwidth]{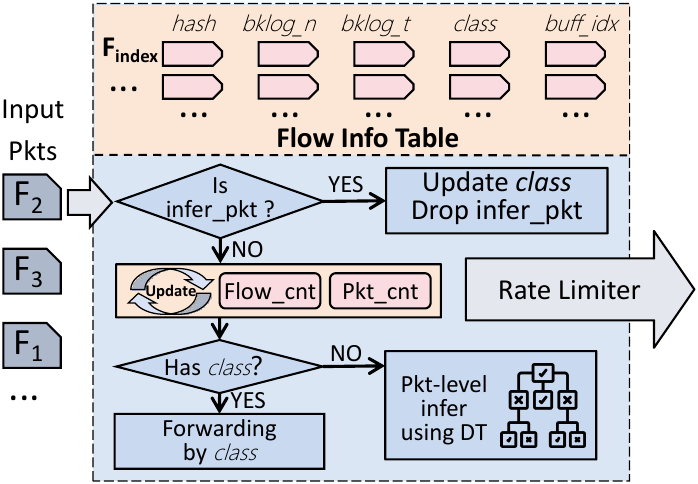}
    \caption{Workflow of Flow Tracker.}
    \label{fig:ft}
    \vspace{-1.5em}
\end{figure}

\vspace{-1em}
\section{Data Engine}
\label{sub:dataengine}
As shown in Figure~\ref{fig:arch}, the Data Engine of \sys is implemented on programmable switch ASICs and consists of three key modules working in concert. The Flow Tracker maintains complex hash table structures to accurately record state information for each network flow, including flow duration $T_i$ and packet count $C_i$, which serve as the basis for subsequent analysis. The Rate Limiter uses an enhanced token bucket algorithm to control the transmission frequency of feature information according to the computed probability function $P(T_i, C_i)$, effectively addressing throughput mismatch between the switch and FPGA, and dynamically adapting to current traffic rate. This probabilistic sampling method, based on flow features, helps maintain system stability under high-traffic conditions while ensuring quality of inputs for model inference. To address FPGA inference latency that could potentially allow some malicious packets to pass undetected in per-packet inference scenarios, \sys integrates lightweight decision trees as a complementary mechanism. For flows awaiting FPGA inference results, the system temporarily uses decision tree inference; once FPGA results become available, the flow processing strategy is immediately updated. The Buffer Manager employs ring buffer techniques to temporarily store feature data, attaches data to mirrored packet headers, and efficiently forwards traffic features to the Model Engine for further processing.

\vspace{-1em}
\subsection{Flow Tracker} 
\noindent\textbf{Overview.}
The Flow Tracker is a critical component of the Data Engine, responsible for tracking network flows and making per-flow decisions. As illustrated in Figure~\ref{fig:ft}, it maintains a Flow Info Table in the switch's Static Random-Access Memory (SRAM), using truncated hash values of five-tuples (source IP, destination IP, source Port, destination Port, and Protocol) as unique flow identifiers. For each flow, the table records several key fields: the flow hash value (\texttt{hash}) for identifying new flow arrivals and handling table collisions; backlog packet count (\texttt{bklog\_n}) and backlog timestamp (\texttt{bklog\_t}) for tracking intervals between feature transmissions; classification results (\texttt{class}) for storing inference outcomes from the Model Engine; and buffer index (\texttt{buff\_idx}), which records the flow’s position in the Buffer Manager’s ring buffer via modulo operation.

Upon packet arrival, the Flow Tracker first determines whether it is an inference packet from Model Engine. If so, it checks whether the packet belongs to a new flow or is the result of a hash collision, and then initializes or updates the corresponding flow entry. For flows with existing classification results, packets are forwarded based on these results. For flows without a classification, a lightweight decision tree implemented on the switch ASIC provides packet-level preliminary inference.

\begin{figure}[t]
    \centering
    \includegraphics[width=0.48\textwidth]{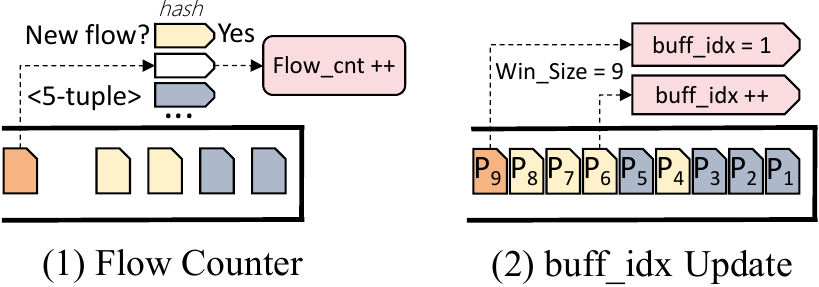}
    \caption{Details in Flow Tracker.}
    \label{fig:fc}
    \vspace{-1.5em}
\end{figure}

\noindent\textbf{Flow Counting Mechanism.}
We define the start time of a flow as the instant when its first packet arrives at the switch, and count the number of flows that send packets within each timeout interval $T_{w}$. This statistical approach, which focuses on newly arrived flows within each interval, helps mitigate bias arising from missing the exact start or end times of flows~\cite{spang2019estimating}. The mechanism operates as illustrated in Figure~\ref{fig:fc} (1): the flow counter detects new flows by checking hash registers associated with each flow. Upon detecting a new flow, the counter increments $n$ by 1. At the end of each $T_{w}$ period, both the hash registers and the flow count $n$ are reset by the control plane to begin a new counting cycle, ensuring accurate and up-to-date flow statistics.

\noindent\textbf{Buffer Index Update.}
As depicted in Figure~\ref{fig:fc} (2), since the data plane cannot perform modulo operations directly, we maintain a separate buffer index (\texttt{buff\_idx}) for each flow that increments with each packet and resets to 1 when reaching buffer size. The flow's packet counter (\texttt{Pkt\_cnt}) independently tracks total packets. This design enables a logical ring buffer where the buffer index cycles through available space, ensuring new data replaces old data in a circular manner.

\begin{figure}[t]
    \centering
    \includegraphics[width=0.48\textwidth]{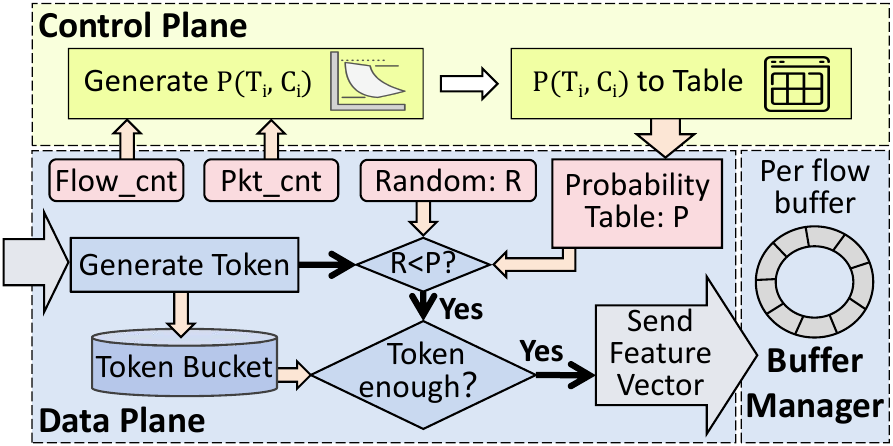}
    \caption{Workflow of Rate Limiter.}
    \label{fig:rl}
    \vspace{-1.5em}
\end{figure}

\vspace{-1em}
\subsection{Rate Limiter}
\noindent\textbf{Overview.}
Figure~\ref{fig:rl} illustrates our novel probabilistic rate limiter design, which integrates a dynamic probability model with the token bucket algorithm to intelligently allocate network traffic resources. This hybrid mechanism is specifically engineered to regulate interactions between the data engine and the model engine, ensuring fair opportunities for flow inference while preventing overload of the model engine.

The rate limiter maintains global traffic statistics within each timing window $T_w$, including \texttt{Flow\_cnt} (the total number of flows) and \texttt{Pkt\_cnt} (the total number of packets processed). While standard work-conserving algorithms such as WFQ (Weighted Fair Queuing) are mature solutions for ensuring inter-flow fairness, our goal is to achieve precise rate control based on per-flow state. In practice, it is infeasible to allocate physical queues for each flow, and alternative approximation schemes cannot guarantee precise rate matching between FPGA and programmable switches. Our probabilistic token bucket mechanism provides a lightweight, hardware-friendly alternative that can be efficiently implemented on existing switch architectures. Leveraging these statistics, the system achieves several advantages: high-speed flows are more likely to fail when requesting tokens, thereby preserving inference opportunities for lower-rate flows, and by capping the token bucket size to no more than the queue length, the system can effectively absorb traffic bursts without causing excessive queuing or packet drops.

\begin{figure}[t]
\centering
\includegraphics[width=\linewidth]{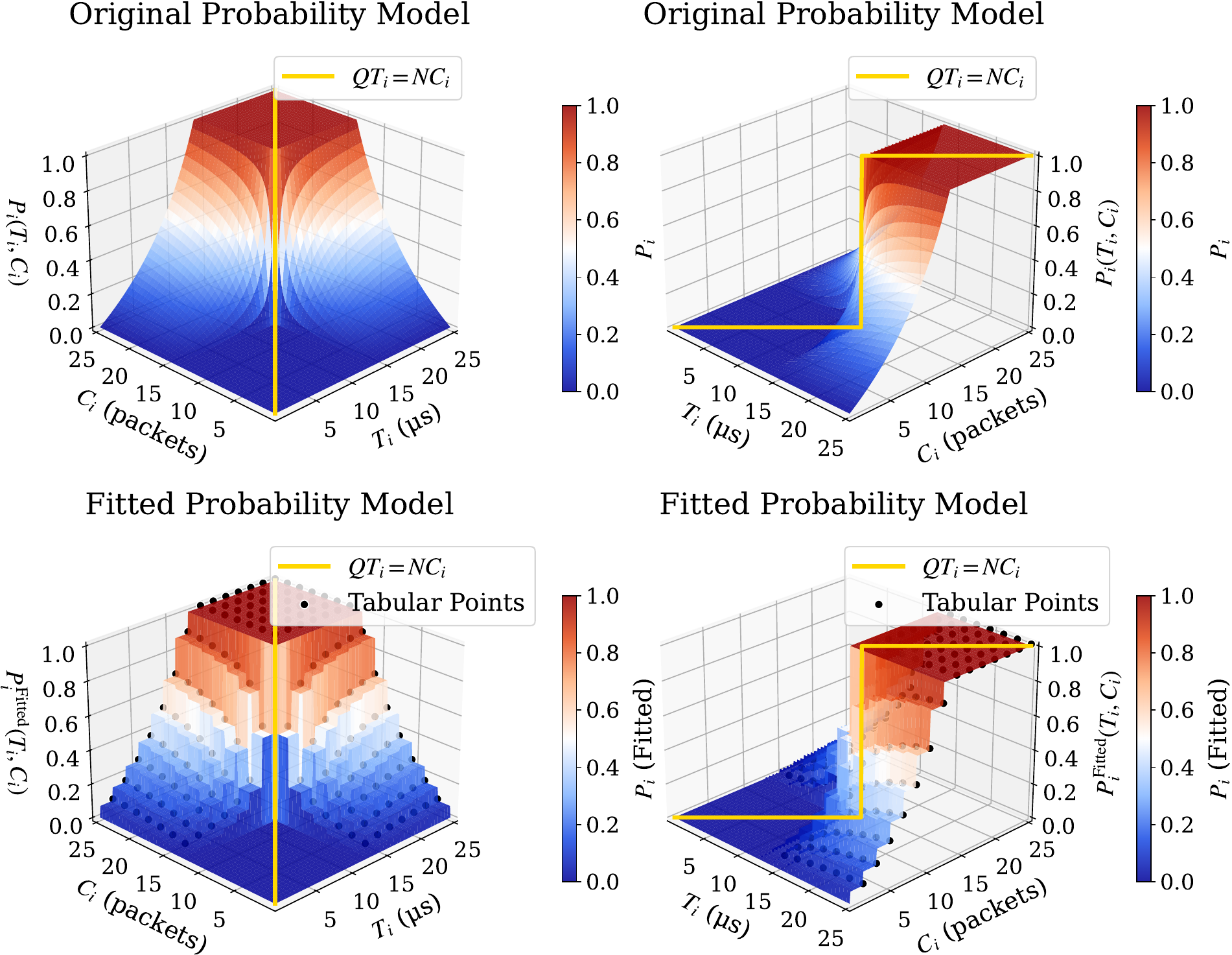}
\caption{Probability curves of token generation model.}
\label{fig:prob_model}
\vspace{-0.5cm}
\end{figure}

\noindent\textbf{Probability-Based Token Allocation Model.}
From a theoretical perspective, the rate limiter constructs a probability function $P(T_i, C_i)$ based on real-time global traffic statistics—specifically, the global flow count $N$ and the aggregate packet rate $Q$. Here, $T_i$ denotes the elapsed time since flow $i$ last transmitted its features, and $C_i$ represents the number of packets transmitted by flow $i$ during this period. By considering both $T_i$ and $C_i$, the system can approximate the instantaneous rate of each flow. For each flow $i$, the token allocation is guided by two key criteria:

\textbf{Criterion 1}: In the idealized scenario where all flows operate at equal rates, each flow should receive tokens at an average interval of $\frac{N}{V}$, where $V$ is the token generation rate.

\textbf{Criterion 2}: In practical situations with heterogeneous flow rates, token allocation should be proportional to each flow’s speed. Thus, the expected token acquisition interval for flow $i$ should approach $\frac{Q}{Q_i V}$, where $Q_i = \frac{C_i}{T_i}$ represents the average packet rate of flow $i$.

The token bucket generation rate $V$ is set according to the communication bandwidth $B$ between engines, FPGA frequency $F$, and feature vector width $W$:
\vspace{-0.5em}
\begin{equation}
V = \min(F, B/W)
\end{equation}

Based on these criteria, we propose the following piecewise probability model.
\vspace{-0.5em}
\begin{equation}
P_i(T_i, C_i) =
\begin{cases}
\frac{C_i(VT_i - N)}{QT_i - NC_i}, & \text{if } T_i \in \left[ \frac{N}{V}, \frac{QT_i}{C_iV} \right] \, \left( \frac{N}{V} < \frac{QT_i}{C_iV} \right) \\
\frac{T_i(VC_i - Q)}{NC_i - QT_i}, & \text{if } T_i \in \left[ \frac{QT_i}{C_iV}, \frac{N}{V} \right] \, \left( \frac{N}{V} > \frac{QT_i}{C_iV} \right) \\
1, & \text{if } QT_i = NC_i \text{ and } T_i \geq \frac{N}{V} \\
0, & \text{if } QT_i = NC_i \text{ and } T_i < \frac{N}{V}
\end{cases}
\end{equation}

\noindent\textbf{Probability Model Deployment.}
To intuitively demonstrate our probability model, Figure~\ref{fig:prob_model} plots the function curves under representative network settings. The illustrated scenario involves 1000 concurrent flows, with the model engine processing packets at 75~Mpps and the network sustaining a total throughput of 1000~Mpps (approximately 800~Gbps, assuming an average packet size of 100 bytes). Given that the data plane cannot directly compute complex probability expressions, the rate limiter discretizes the model in the control plane by constructing a lookup table: the ranges of $T_i$ and $C_i$ are uniformly partitioned, enabling efficient mapping of all possible value pairs to probabilities in the range $[0, 1]$. Figure~\ref{fig:prob_model} presents both the original probability function and its table-based approximation, illustrating that our implementation closely preserves the intended behavior of the model.

\begin{algorithm}[t] 
	\caption{Token Bucket Algorithm for Rate Limiter}
	\label{algorithm:tb}
	\begin{algorithmic}[1]
	    \If {$T_{last} = 0$}
	        \State $T_{last} \leftarrow T_{now}$, $gap \leftarrow 0$ \Comment{{\footnotesize\textcolor{comment}{Initialize for first packet}}}
	    \Else
	        \State $gap \leftarrow T_{now} - T_{last}$, $T_{last} \leftarrow T_{now}$ \Comment{{\footnotesize\textcolor{comment}{Calculate time gap}}}
	    \EndIf
	    
	    \State $rand \leftarrow$ Random(), $prob \leftarrow$ LookupProbability()
	    \State $bucket \leftarrow bucket + gap$ \Comment{{\footnotesize\textcolor{comment}{Refill tokens}}}
	    
	    \If {$rand < prob$} \Comment{{\footnotesize\textcolor{comment}{Selected for sampling}}}
	        \If {$bucket \geq cost$}
	            \State $bucket \leftarrow bucket - cost$ \Comment{{\footnotesize\textcolor{comment}{Consume token}}}
	            \State SendFeatureVector()
	        \EndIf
	    \EndIf
        
	\end{algorithmic}

\end{algorithm}

\noindent\textbf{Token Bucket Algorithm Implementation.}
In our data plane design, the rate limiter tracks three essential state variables within a custom timing window $T_w$: the number of active flows, the total packet count, and the last packet arrival time $T_{last}$. The details of our probabilistic token bucket mechanism are presented in Algorithm~\ref{algorithm:tb}. Upon the arrival of each packet, the algorithm first computes the time interval between the current arrival time $T_{now}$ and $T_{last}$ to determine the appropriate token replenishment (lines 1--4). Next, a random number $rand$ is generated and compared against the probability value $prob$ retrieved from the precomputed lookup table (line 5). If the packet is probabilistically selected ($rand < prob$) and the token bucket contains enough tokens, the required token $cost$ is deducted from the bucket and the feature vector is transmitted (lines 6--8). Otherwise, if the selection fails or tokens are insufficient, the algorithm only updates the token count based on the elapsed $gap$ (lines 9--14).

\noindent\textbf{Discussion.} The Rate Limiter offers several notable advantages. By probabilistically denying token requests from high-speed flows, it helps reserve transmission opportunities for slower flows, promoting fairness across diverse traffic patterns. Additionally, by capping the token bucket capacity to not exceed the queue length, the system can efficiently accommodate bursty transmissions without risking buffer overflow. Overall, this module enables more equitable and adaptable traffic management, while maintaining stability and robustness within the system.

\vspace{-1em}
\subsection{Buffer Manager}
\vspace{-0.5em}
\noindent\textbf{Ring Buffer Design.}
Inspired by BoS~\cite{yan2024brain}, the Buffer Manager assigns a dedicated ring buffer to each network flow, which temporarily stores feature vectors until they are processed by the Model Engine. For each flow, the ring buffer holds features extracted from preceding packets ($F_1$--$F_8$), including packet intervals and packet lengths, as well as the current packet's feature ($F_9$, stored in metadata). This architecture allows the system to preserve the temporal and sequential relationships between packets within the same flow, offering a comprehensive context for subsequent analysis. When a buffer reaches capacity, it operates in a circular (FIFO) manner, overwriting the oldest features with new entries. This ensures that the buffer consistently reflects the most recent traffic characteristics, efficiently utilizes memory, and maintains the necessary flow context for downstream processing.

\begin{figure}[t]
    \centering
    \includegraphics[width=0.45\textwidth]{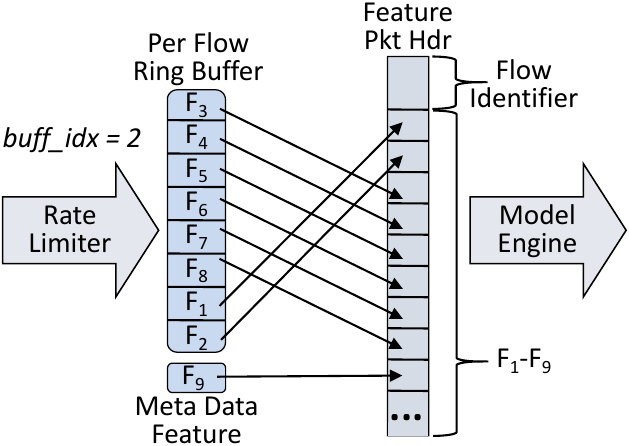}
    \caption{Mechanism of Buffer Manager.}
    \label{fig:bm}
    \vspace{-1.5em}
\end{figure}

\noindent\textbf{Feature Transfer Process.}
The Buffer Manager operates in close coordination with the Rate Limiter to manage feature transmission. As illustrated in Figure~\ref{fig:bm}, when the Rate Limiter determines that feature information should be exported, the Buffer Manager reads the \texttt{buff\_idx} value from the Flow Tracker (e.g., \texttt{buff\_idx}=2 as shown in the figure) and extracts the corresponding feature vectors from the flow's ring buffer in SRAM, assembling them into the packet header. The newest feature, stored in metadata, is appended to the end of this header. During the final Deparser Stage, the Buffer Manager inserts the constructed header into a mirrored packet. These mirrored packets encapsulate both the flow’s five-tuple identification and the sequence of feature vectors, providing the Model Engine with the necessary context for processing.

\vspace{-1em}
\section{Model Engine}
\begin{figure}[t]
    \centering
    \includegraphics[width=0.50\textwidth]{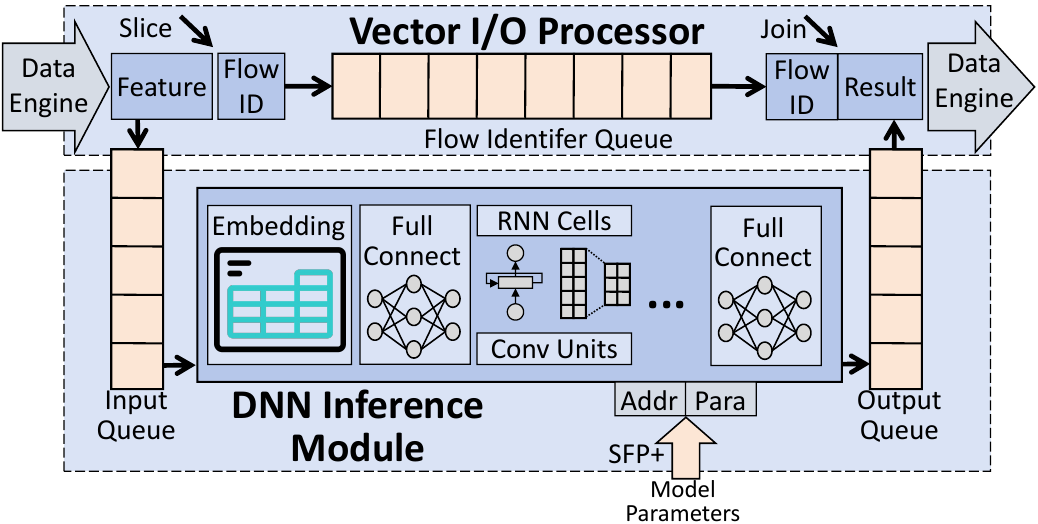}
    \caption{Workflow of Model Engine.}
    \label{fig:me}
    \vspace{-0.5cm}
\end{figure}

\vspace{-0.5em}
\label{sub:modelengine}
The Model Engine performs low-latency inference on traffic features from the Data Engine and is implemented on an FPGA. As shown in Figure~\ref{fig:me}, it comprises two components: a Vector I/O Processor which manages flow-info and a DNN Inference Module which executes model inference. Upon receiving packets, the Vector I/O Processor separates each into flow identifier and feature vector. Flow identifiers are stored in the Flow Identifier Queue to preserve ordering, while feature vectors are forwarded to DNN Inference Module.

 After inference completion, the results are placed in the Output Queue. Each inference result in the Output Queue is paired with the corresponding flow identifier retrieved from the front of the Flow Identifier Queue, forming a new packet with both fields. This packet is then transmitted back to the Data Engine, enabling the switch chip to execute follow-up operations on the relevant flow based on the inference result. This approach maintains correct mapping between flows and their inference results throughout processing.

\vspace{-1em}
\subsection{Vector I/O Processor}
The Vector I/O Processor is responsible for parsing incoming network packets to extract both the flow identifier (such as the five-tuple) and the associated feature vector. To maintain the correspondence between flows and their features throughout the inference process, the processor employs a FIFO queue to store flow identifiers. Meanwhile, the DNN Inference Module utilizes asynchronous FIFO queues for both its input and output, enabling seamless data transfer between modules operating under different clock domains. This asynchronous FIFO design not only decouples the timing dependencies between the Vector I/O Processor and the DNN Inference Module, but also improves system robustness and throughput.

During operation, the Vector I/O Processor continuously monitors the status of both the flow identifier FIFO and the output FIFO of the DNN Inference Module. Whenever both queues contain valid entries, the processor simultaneously dequeues the head elements, assembles the flow identifier and inference result into a new packet, and transmits this packet to the Data Engine. The Data Engine then forwards the result to the programmable switch, ensuring timely and accurate flow-level actions based on the inference outcome. For example, when a flow is identified as malicious based on the inference result, the programmable switch can record this status in its flow register. Subsequently, whenever packets belonging to this flow are encountered, the switch can enforce corresponding actions such as rate limiting or traffic isolation, thereby mitigating potential security threats in real time.

\subsection{DNN Inference Module} 
The DNN Inference Module performs neural network inference for traffic analysis directly on the FPGA, as illustrated in Figure~\ref{fig:me}. The core computation uses a systolic array optimized for INT8 operations, enabling efficient matrix-vector multiplications for several common neural network layers. Our implementation supports embedding lookups mapped to LUTs, fully connected (FC) layers, convolution (Conv) layers, and recurrent layers. These layers are composed sequentially according to the model architecture, with all layer parameters and feature dimensions fixed at synthesis time to match application requirements and FPGA resource constraints. Although the implementation is not fully modular, each layer's computation is mapped onto the same systolic array, and the data path is managed to support the required layer ordering.

Model parameters, including weights and biases for all supported layers, are loaded from the host into on-chip memory via the network interface. During inference, feature vectors are processed in batches: embedding lookups are performed first, followed by the configured sequence of FC, Conv, and recurrent layers, all using offline-quantized INT8 arithmetic. Asynchronous FIFO queues decouple dataflow between layers and enable efficient pipelining. This design provides low-latency inference for network data while making efficient use of FPGA compute and memory resources.
\vspace{-1em}
\section{Implementation}
\label{sec:implementation}
\begin{figure}[t]
    \centering
    \includegraphics[width=0.48\textwidth]{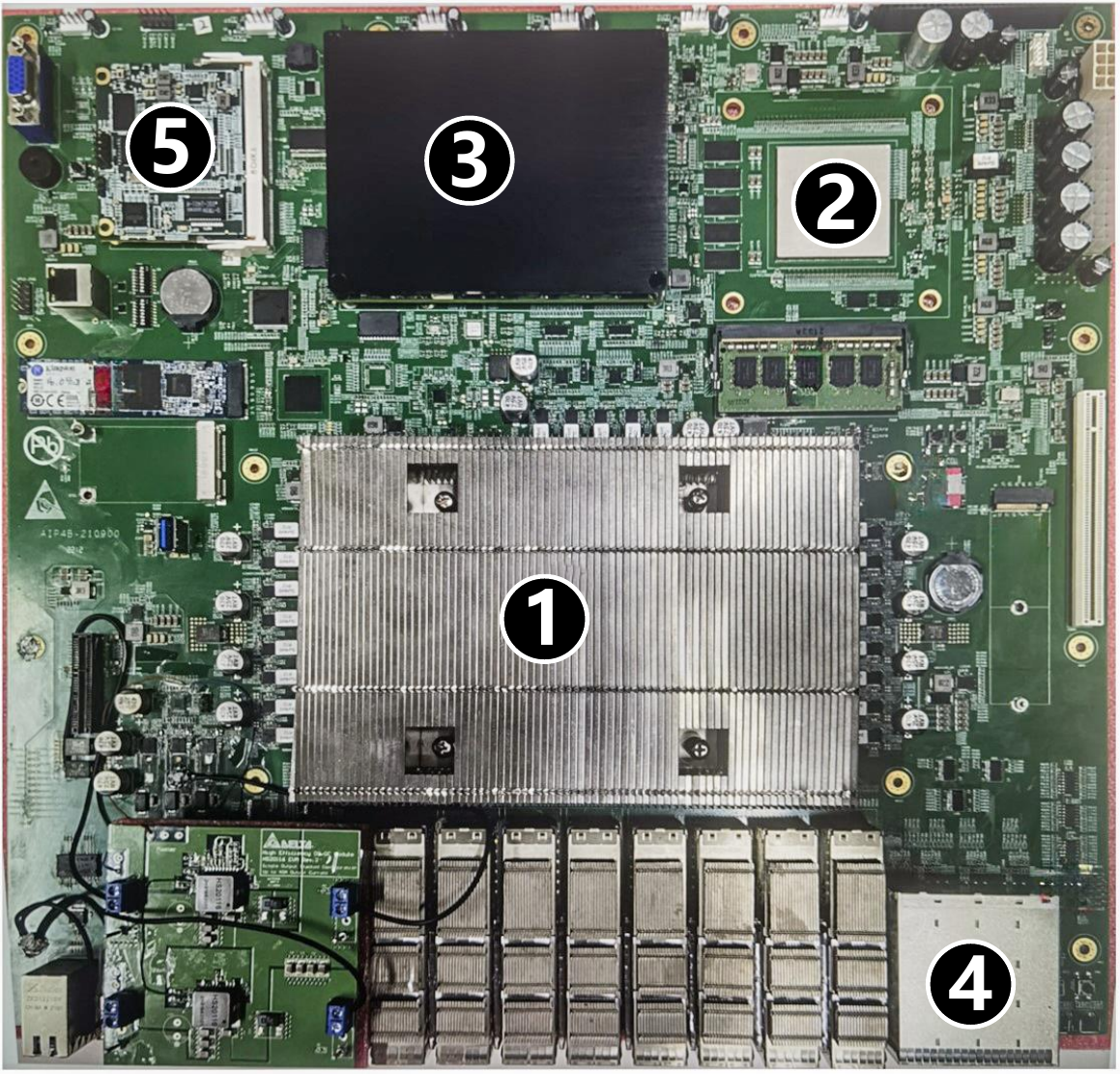}
    \caption{Hardware Platform of \sys: The FPGA-Enhanced Programmable Switch Implementation.}
    \label{fig:realpcb}
    \vspace{-0.5cm}
\end{figure}

\vspace{-0.5em}
\noindent\textbf{Experimental Hardware.} As shown in Figure~\ref{fig:realpcb}, our experiments are conducted on a high-performance programmable switch platform that integrates both FPGA and Tofino chips on a single board. The hardware is implemented using a 22-layer high-performance printed circuit board (PCB), utilizing back-drilling technology to balance 100~Gbps high-speed interconnection performance and manufacturing costs. The switching system is equipped with a dedicated power control network, supporting a core operating current of up to 100~A and peak control currents up to 300~A, with precise impedance control to ensure reliable power delivery. The overall design process included approximately half a month for requirement analysis, one month for schematic design, two months for PCB design, and one and a half months for PCB manufacturing, soldering, and prototype hardware debugging. The interface design and testing between FPGA and Tofino required an additional month, with software driver development and system-level integration carried out in parallel. 

Figure~\ref{fig:realpcb} presents a photograph of the hardware prototype, with the primary components highlighted as follows:
\begin{itemize}[itemsep=0pt]
    \item[\blackcirclednum{1}] \textbf{Tofino 2 switch chip}~\cite{tofino2}: Featuring 20 MAU stages, 200~Mbits of SRAM, and 10.3~Mbits of TCAM per pipeline. We configured this chip to efficiently implement our Data Engine modules while balancing resource utilization across pipeline stages. Multiple 100~Gbps port channels were designed to connect the Tofino chip to the FPGA, enabling high-bandwidth data exchange.
    \item[\blackcirclednum{2}] \textbf{Xilinx ZU19EG FPGA chip}~\cite{amd_zynq_ultrascale_plus_mpsoc}: Integrates approximately 80~Mbits of on-chip memory and 1,143,450 logic elements. The FPGA resources were partitioned to accommodate the Model Engine's Vector I/O Processor and Neural Computing Array, while maintaining timing closure at high clock frequencies. The FPGA is directly connected to the Tofino chip through high-speed port channels, enabling low-latency communication.
    \item[\blackcirclednum{3}] \textbf{Control plane CPU}: Programmed to manage the Tofino chip, responsible for configuration, monitoring, and probability model adjustments.
    \item[\blackcirclednum{4}] \textbf{Four front-panel FPGA ports}: Designed with configurable transceivers to operate in either 10G or 25G modes for external communication. The design process required careful signal integrity analysis and impedance matching.
    \item[\blackcirclednum{5}] \textbf{BMC module}: Customized for hardware status monitoring and remote management of the platform, including thermal management for high-performance components.
\end{itemize}

\noindent\textbf{Experimental Software.}
Our development process encompassed multiple hardware and software platforms, requiring the integration of a diverse set of technologies and toolchains. For the Tofino chip, we used SDE version 9.8.0 for software development and debugging, while FPGA programming was conducted using Vivado and Vitis HLS 2024.1~\cite{AMD2024VitisHLS}. The prototype system comprises: (1) approximately 1,500 lines of P4 code~\cite{bosshart_p4_2014} to implement the Data Engine's three core modules—Flow Tracker, Rate Limiter, and Buffer Manager—on the programmable switch ASICs, which involved precise implementation of hash table maintenance, token bucket algorithms, and feature cache rings; (2) 1,000 lines of HLS code for the Model Engine's two main components—Vector I/O Processor and DNN Inference Module—on the FPGA, responsible for parallel operations and efficient matrix processing; (3) 1,500 lines of Python code and 1,000 lines of C++ code for model training, testing, and quantization, ensuring efficient model execution under the resource constraints of the hardware; and (4) 300 lines of Python code for calculating probability models in the control plane, enabling intelligent feature transmission decisions and adaptive flow management. Additionally, we developed an 800-line Python simulation program to evaluate and validate \sys's performance and scalability under high-throughput and large-scale network scenarios.

\noindent\textbf{Model Training and Quantization.}
We trained CNNs and RNNs for traffic classification using protocol-agnostic features: packet lengths and inter-packet arrival times. These features capture temporal patterns without requiring protocol-specific information. Our models include normalization layers, convolutional/recurrent units for temporal dependencies, and fully connected layers with ReLU activation. To address class imbalance, we applied oversampling and undersampling techniques during preprocessing.

After training, we used Vitis-AI~\cite{AMD2023VitisAI} to quantize models from floating-point to INT-8 format for efficient FPGA deployment. The quantization process assigns different decimal positions to layers based on activation distributions to preserve accuracy. Our evaluation shows that quantization substantially reduces computation and storage requirements while maintaining classification performance with negligible degradation, enabling high-throughput, low-latency inference on resource-constrained FPGA platforms.
\vspace{-0.5cm}
\section{EVALUATION}
\vspace{-0.2cm}

\label{sec:evaluation}
Our evaluation addresses several key questions: (i) What classification accuracy can \sys achieve on real traffic datasets, and how does it compare to other methods? (ii) What computational resources are required by \sys, including both programmable switching ASICs during feature extraction and FPGA resources? (iii) How is \sys performance affected under different scales of concurrent flows and throughput? (iv) As an asynchronous hybrid system, does \sys introduce noticeable latency overhead when using FPGA for inference?

To answer these questions, we implemented \sys and other baselines on an FPGA-enhanced programmable switch using P4\cite{bosshart_p4_2014} and HLS~\cite{AMD2024VitisHLS} (Details in Section \S~\ref{sec:implementation}), and conducted comparative evaluations. We used publicly available datasets for VPN encrypted traffic and malware detection for testing.

\vspace{-1em}
\subsection{Methodology}

\noindent\textbf{Testbed Setup.} We implemented \sys on an FPGA-enhanced programmable switch (Details in Section \S~\ref{sec:implementation}) using P4 and HLS, connected to a high-performance server equipped with dual Intel Xeon 5418Y processors and 512GB of memory. The server is equipped with two NVIDIA MCX75310AAS-NEAT 400G network interface cards, with one dedicated to traffic generation and the other to packet reception. The sending NIC generates synthetic traffic and replays pcap files using DPDK (v23.11.3) pktgen~\cite{dpdk_2025}, while the receiving NIC collects and analyzes the processed packets using the DPDK environment.

\noindent\textbf{Schemes compared.} We compared \sys with the following 9 schemes:

\noindent\textbf{(a) \sys-CNN-FLOW:} We implemented a flow-level CNN with 3 convolutional layers and 2 fully connected layers. Flow accuracy is calculated based on majority voting of packet results within each flow.

\noindent\textbf{(b) \sys-RNN-FLOW:} We developed a flow-level RNN with a single custom RNN cell and a dense output layer. The model processes packet length and IPD features through embeddings and classifies flows using the final hidden state.

\noindent\textbf{(c) FlowLens~\cite{DBLP:conf/ndss/flowlens}:} We implemented packet statistics in Tofino using FlowLens' FMA code with control plane inference. Our XGBoost model used default parameters\cite{xgboost_documentation_2025}, with collection windows adjusted based on throughput.

\noindent\textbf{(d) \sys-CNN-PKT:} We created a packet-level CNN with the same network architecture as \sys-CNN-FLOW. This model processes individual packets independently, with accuracy calculated on per-packet classification results.

\noindent\textbf{(e) \sys-RNN-PKT:} We implemented a packet-based RNN with the same network architecture as \sys-RNN-FLOW. Accuracy is measured on a per-packet basis, evaluating the model's ability to correctly classify each individual packet.

\noindent\textbf{(f) Netbeacon~\cite{DBLP:conf/uss/netbeacon}:} We implemented multi-phase tree models. Each phase uses a Random Forest (3 trees, depth 7) matching their configuration.

\noindent\textbf{(g) Leo~\cite{jafri_leo_2024}:} We implemented a decision tree (max depth 22, up to 1024 leaf nodes) on switches using packet length extremes and cumulative flow length.

\noindent\textbf{(h) BoS~\cite{yan2024brain}:} We implemented the largest variant with a binarized GRU network (9-bit hidden states), 8 GRU units, 6-bit embeddings, and complete feature embedding-GRU-output architecture.

\noindent\textbf{(i) N3IC~\cite{siracusano_re-architecting_2022}:} We implemented a binary MLP on SmartNIC using flow-level and packet-level features with hidden layers [128, 64, 10]. Due to hardware constraints, we simulated switch-side logic and inference in software.

\noindent\textbf{Tasks.} We use the following tasks to evaluate \sys, as summarized in Table~\ref{tab:dataset_transposed}. 
(i) Encrypted Traffic Classification on VPN: This task aims to classify network traffic that has been encrypted by VPNs. We use the ISCXVPN2016 dataset~\cite{iscxvpn}.
(ii) Malware Identification: This task distinguishes between traffic generated by benign applications and various types of malware. We use the USTC-TFC2016 dataset~\cite{ustc}. Note that model training is performed offline, and the FPGA is used exclusively for inference tasks.

\noindent\textbf{Metrics.} We use flow-level macro-F1 as the accuracy metric for FlowLens, while for other schemes we report packet-level macro-F1. We also provide a breakdown of Precision and Recall for each class to enable detailed performance analysis.

\renewcommand{\arraystretch}{1.2}

\begin{table}[t]
  \centering
  \caption{Experimental settings.}
  \label{tab:dataset_transposed}
  \resizebox{\linewidth}{!}{%
  \begin{tabular}{lcc}
    \toprule
    \textbf{Dataset (Task)} & \textbf{ISCXVPN2016}~\cite{iscxvpn} & \textbf{USTC-TFC}~\cite{ustc} \\
    \midrule
    \textbf{Training Flows}      & 29,295  & 101,789 \\
    \textbf{Test Flows}          & 7,328   & 25,455 \\
    \textbf{Number of Classes}   & 7       & 12     \\
    \textbf{Class Ratio}         & 
    \renewcommand{\arraystretch}{0.8}%
    \begin{tabular}[c]{@{}c@{}}11:4:13:10:\\18:128:1\end{tabular}
    & 
    \renewcommand{\arraystretch}{0.8}%
    \begin{tabular}[c]{@{}c@{}}92:10:4:14:17:23:\\105:1:16:132:27:1\end{tabular} \\
    \textbf{Optimizer}           & AdamW   & AdamW  \\
    \textbf{Learning Rate}       & 0.01    & 0.005  \\
    \bottomrule
  \end{tabular}%
  }
  \vspace{-0.7cm}
\end{table}

\begin{table*}[t]
\centering
\caption{Performance Comparison of Different Methods on Encrypted Traffic Classification and Malware Detection}
\label{tab:acc}
\renewcommand{\arraystretch}{1.1}
\footnotesize
\resizebox{\textwidth}{!}{
\begin{tabular}{l|c|c|c||c|c|c|c|c|c}
\toprule
\textbf{Class} 
& \textbf{\sys\textsubscript{F-CNN}} 
& \textbf{\sys\textsubscript{F-RNN}} 
& \textbf{FlowLens~\cite{DBLP:conf/ndss/flowlens}} 
& \textbf{\sys\textsubscript{P-CNN}} 
& \textbf{\sys\textsubscript{P-RNN}} 
& \textbf{NetBeacon~\cite{DBLP:conf/uss/netbeacon}} 
& \textbf{Leo~\cite{jafri_leo_2024}} 
& \textbf{BoS~\cite{yan2024brain}} 
& \textbf{N3IC~\cite{siracusano_re-architecting_2022}} \\
\midrule
\multicolumn{10}{c}{\textbf{Encrypted Traffic Classification (ISCXVPN2016~\cite{iscxvpn})}} \\
\midrule
Chat      & 0.883/0.852 & 0.939/0.882 & 0.862/0.922 & 0.917/0.836 & 0.804/0.653 & 0.627/0.169 & 0.489/0.353 & 0.922/0.913 & 0.533/0.527 \\
Email     & 0.924/0.834 & 0.944/0.924 & 0.888/0.821 & 0.862/0.882 & 0.893/0.746 & 0.230/0.321 & 0.333/0.019 & 0.932/0.925 & 0.349/0.353 \\
File      & 0.879/0.849 & 0.923/0.912 & 0.860/0.889 & 0.886/0.797 & 0.889/0.844 & 0.861/0.701 & 0.815/0.720 & 0.915/0.922 & 0.820/0.848 \\
P2P       & 0.977/0.963 & 0.976/0.988 & 0.923/0.913 & 0.911/0.914 & 0.932/0.947 & 0.861/0.908 & 0.853/0.867 & 0.925/0.917 & 0.905/0.892 \\
Stream    & 0.902/0.968 & 0.919/0.973 & 0.966/0.959 & 0.877/0.965 & 0.894/0.969 & 0.890/0.976 & 0.850/0.944 & 0.916/0.908 & 0.886/0.938 \\
Voip      & 0.989/0.995 & 0.992/0.996 & 0.998/0.995 & 0.999/0.998 & 0.999/0.999 & 0.986/0.993 & 0.994/0.997 & 0.829/0.723 & 0.994/0.993 \\
Web       & 0.803/0.662 & 0.793/0.625 & 0.700/0.525 & 0.800/0.861 & 0.869/0.821 & 0.860/0.405 & 0.784/0.046 & 0.729/0.623 & 0.856/0.524 \\
\midrule
Macro-F1 & 0.890 & 0.912 & 0.870 & 0.892 & 0.873 & 0.658 & 0.578 & 0.863 & 0.738 \\
\midrule
\multicolumn{10}{c}{\textbf{Malware Detection (USTC-TFC~\cite{ustc})}} \\
\midrule
Cridex     & 0.999/1.000 & 0.999/1.000 & 0.999/0.998 & 0.999/1.000 & 0.996/1.000 & 0.933/0.997 & 0.984/0.995 & 0.983/0.996 & 0.866/0.861 \\
FTP        & 0.999/0.998 & 0.999/0.998 & 0.999/0.998 & 0.997/1.000 & 0.993/1.000 & 0.993/0.485 & 0.928/0.986 & 0.986/0.791 & 0.848/0.853 \\
Geodo      & 0.979/0.881 & 0.984/0.891 & 0.945/0.905 & 0.932/0.343 & 0.786/0.424 & 0.899/0.524 & 0.369/0.188 & 0.934/0.615 & 0.857/0.851 \\
Htbot      & 0.962/0.987 & 0.959/0.989 & 0.964/0.984 & 0.932/0.986 & 0.938/0.982 & 0.830/0.782 & 0.897/0.935 & 0.919/0.937 & 0.871/0.867 \\
Neris      & 0.921/0.594 & 0.888/0.732 & 0.902/0.817 & 0.766/0.473 & 0.736/0.575 & 0.665/0.469 & 0.584/0.453 & 0.761/0.610 & 0.852/0.847 \\
Nsis-ay    & 0.985/0.970 & 0.971/0.982 & 0.985/0.986 & 0.959/0.968 & 0.962/0.972 & 0.984/0.912 & 0.918/0.960 & 0.963/0.968 & 0.860/0.855 \\
Warcraft   & 0.998/0.999 & 0.996/0.994 & 0.995/0.993 & 1.000/1.000 & 0.999/1.000 & 0.890/1.000 & 0.995/0.997 & 0.956/0.998 & 0.855/0.859 \\
Zeus       & 0.932/0.945 & 0.955/0.863 & 0.971/0.932 & 0.962/0.978 & 0.978/0.958 & 0.823/0.260 & 0.895/0.791 & 0.976/0.951 & 0.865/0.860 \\
Virut      & 0.671/0.941 & 0.760/0.893 & 0.827/0.908 & 0.723/0.863 & 0.763/0.836 & 0.571/0.815 & 0.691/0.692 & 0.744/0.844 & 0.859/0.855 \\
Weibo      & 0.684/0.822 & 0.702/0.927 & 0.745/0.804 & 0.655/0.821 & 0.685/0.817 & 0.649/0.789 & 0.653/0.708 & 0.652/0.837 & 0.862/0.859 \\
Shifu      & 0.999/0.966 & 0.997/0.995 & 0.982/0.914 & 0.947/0.829 & 0.965/0.767 & 0.194/0.302 & 0.610/0.593 & 0.835/0.501 & 0.862/0.859 \\
SMB        & 0.700/0.522 & 0.845/0.504 & 0.726/0.654 & 0.632/0.416 & 0.665/0.492 & 0.607/0.434 & 0.555/0.491 & 0.651/0.406 & 0.862/0.859 \\
\midrule
Macro-F1 & 0.887 & 0.901 & 0.914 & 0.907 & 0.838 & 0.670 & 0.741 & 0.814 & 0.858 \\
\bottomrule
\end{tabular}
}
\vspace{0.3em}
\caption*{\footnotesize
\textbf{Note:} The four \sys~columns represent models using flow-level statistics with CNN (\sys\textsubscript{F-CNN}) and RNN (\sys\textsubscript{F-RNN}), as well as packet-level statistics with CNN (\sys\textsubscript{P-CNN}) and RNN (\sys\textsubscript{P-RNN}).
}
\vspace{-1cm}
\end{table*}

\vspace{-1em}
\subsection{Classification Accuracy on Real Datasets}
\vspace{-0.5em}
\label{sec:eva_accuracy}
Table \ref{tab:acc} presents a comprehensive comparison of accuracy and recall performance of different methods on encrypted traffic classification and malware detection tasks, evaluated at both the flow and packet levels. Across all categories in both tasks, \sys consistently demonstrates superior accuracy compared to other baselines. At the flow level, the macro-F1 score of \sys reaches 0.890, which is very close to that of FlowLens (0.870), indicating that \sys can achieve state-of-the-art detection performance while retaining the advantages of a neural network-based approach.

When evaluated at the packet level, \sys outperforms all other compared methods across both tasks. On the malware detection benchmark, \sys achieves a macro-F1 score of 0.907, greatly surpassing tree-based approaches such as NetBeacon (0.670) and showing a clear margin over advanced neural baselines like N3IC (0.858). Compared to other neural models, including Leo (0.741) and BoS (0.814), \sys also demonstrates consistently higher accuracy.

This performance gap becomes especially pronounced in challenging multiclass scenarios, where traditional methods often suffer from error propagation or limited packet-level representational capacity. While BoS and Leo achieve moderate results, they still lag behind \sys, highlighting the benefits of high-precision inference enabled by the \sys architecture. Notably, BoS is constrained by model binarization and the limited size of each component that switching ASICs can accommodate, which reduces its accuracy. Tree-based methods like NetBeacon can only update predictions at discrete points, further limiting performance on fine-grained, packet-level tasks. Although \sys does not perform continuous inference for every packet, its higher per-inference accuracy ensures better overall packet-level accuracy. This advantage is particularly evident in complex multiclass settings, where the CNN-based \sys model achieves the highest classification performance among all tested methods.

\vspace{-1em}
\subsection{Hardware Resource Utilization}
\vspace{-0.5em}
\label{sec:eva_resources}
Table~\ref{tab:p4_resource} compares the hardware resource overhead of several representative P4 systems on programmable switches. The results show that \sys achieves relatively low resource usage across multiple dimensions. For instance, its SRAM consumption (12.9\%) is much lower than FlowLens (34.2\%), indicating that the introduction of FPGA effectively reduces the resource overhead of P4 systems. In terms of TCAM usage, \sys also stands out, with only 4.4\% overhead compared to NetBeacon’s 18.8\%, highlighting that \sys avoids the trade-off of reducing one resource at the cost of sharply increasing another. Beyond memory, \sys maintains efficient use of bus bandwidth and pipeline stages, remaining below or comparable to most other systems. This balanced and predictable resource profile is enabled by the decoupled DataEngine design, which keeps \sys’s overhead stable across different tasks and workloads.

Table~\ref{tab:resource_summary} summarizes the FPGA resource utilization, where percentages represent the proportion of total available resources consumed by each module. The results show that LUTs and FFs are the primary resources consumed, particularly in the core computational components of the CNN and RNN modules. For example, the overall CNN module uses 38.4\% of LUTs and 33.8\% of FFs. While BRAM and DSP usage is higher in the overall modules (up to 7.1\% and 8.1\% for CNN, and 6.3\% and 4.6\% for RNN, respectively), it remains minimal within individual submodules, where most values are below 4\%. This indicates that our design primarily leverages combinational and sequential logic to efficiently implement neural network inference on the FPGA, while memory and DSP consumption remains moderate. Overall, the resource utilization is low, leaving ample headroom for further optimization and larger deployments.
\begin{table}[t]
  \centering
  \caption{P4 Systems Resource Overhead Comparison}
  \renewcommand{\arraystretch}{1.1}
  \footnotesize
  \resizebox{\linewidth}{!}{
    \begin{tabular}{l|c|c|c|c}
      \toprule
      \textbf{System} & \textbf{SRAM} & \textbf{TCAM} & \textbf{Bus} & \textbf{Stage} \\
      \midrule
      \sys      & 12.9\% & 4.4\% & 3.5\% & 9 \\
      FlowLens~\cite{DBLP:conf/ndss/flowlens}     & 34.2\% & 0.0\% & 2.4\% & 9 \\
      BoS~\cite{yan2024brain}         & 26.3\% & 6.3\% & 8.6\% & 12 \\
      Leo~\cite{jafri_leo_2024}       & 26.9\% & 9.0\% & 5.2\% & 12 \\
      NetBeacon~\cite{DBLP:conf/uss/netbeacon}    & 11.6\% & 18.8\% & 6.4\% & 12 \\
      \bottomrule
    \end{tabular}
  }
  \vspace{-0.7cm}
  \label{tab:p4_resource}
\end{table}
\begin{table}[t]
  \centering
  \caption{Neural Network Resource Utilization (\%)}
  \label{tab:resource_summary}
  \renewcommand{\arraystretch}{1.1}
  \footnotesize
  \resizebox{\linewidth}{!}{
    \begin{tabular}{l|cccc}
      \toprule
      \textbf{Module} & \textbf{LUT} & \textbf{FF} & \textbf{BRAM} & \textbf{DSP} \\
      \midrule
      CNN (overall)         & 38.4\% & 33.8\% &  7.1\% &  8.1\% \\
      \quad Embedding       &  4.2\% &  5.1\% &  0.5\% &  0.0\% \\
      \quad Convolutional   & 25.6\% & 19.7\% &  4.0\% &  5.7\% \\
      \quad FC              &  8.6\% &  9.0\% &  2.6\% &  2.4\% \\
      RNN (overall)         & 25.6\% & 31.2\% &  6.3\% &  4.6\% \\
      \quad Embedding       &  4.2\% &  5.1\% &  0.5\% &  0.0\% \\
      \quad Recurrent       & 15.8\% & 18.7\% &  3.6\% &  2.4\% \\
      \quad FC              &  8.6\% &  9.2\% &  2.2\% &  2.2\% \\
      Vector I/O            &  6.0\% &  4.8\% &  0.3\% &  0.0\% \\
      \bottomrule
    \end{tabular}
  }\vspace{-0.3cm}
\end{table}
\begin{figure*}[ht]
    \centering
    \includegraphics[width=0.96\linewidth]{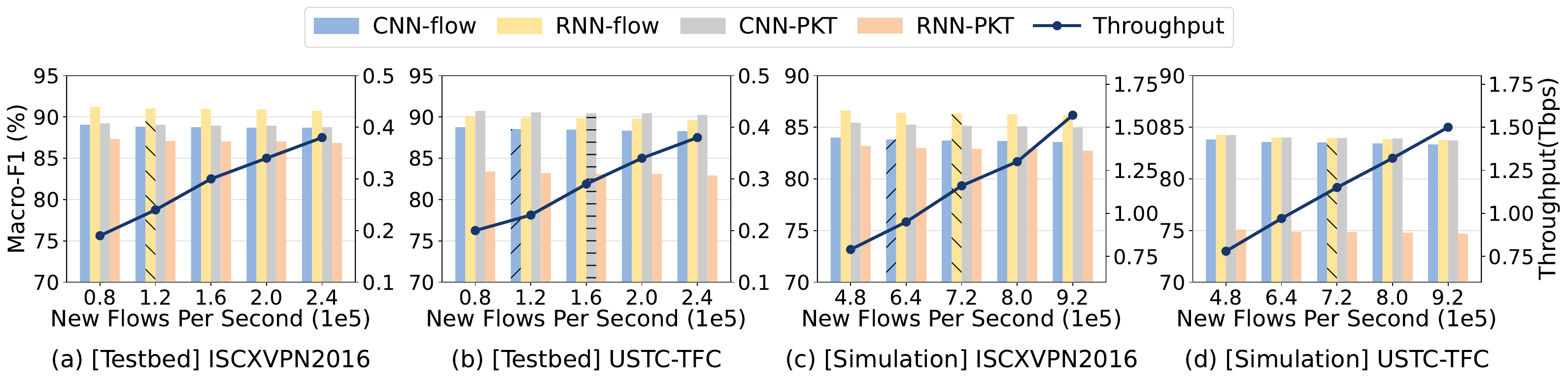}
    \caption{Scaling test of \sys}
    \label{fig:scale}
    \vspace{-0.5cm}
\end{figure*}
\vspace{-1em}
\subsection{Flow Count and Throughput Scalability}
\vspace{-0.5em}
\label{sec:eva_scalability}

We conduct stress tests on \sys under high-concurrency and high-throughput scenarios. Since original network trace files were collected in low-bandwidth networks, we generate high-throughput trace files by concurrently packaging flows with unique identifiers and reassigning timestamps. Figure \ref{fig:scale} shows results where we progressively increase flow concurrency until the traffic generator's NIC bandwidth is saturated. \sys easily handles this scale, with macro-F1 scores nearly identical to Table \ref{tab:acc}.

To evaluate larger scales, we build a simulator emulating \sys's workflow. We configure maximum concurrent flow count based on switch chip registers and set inference latency according to measured results. After validating simulator accuracy against our testbed, we explore significantly larger scales with tens of thousands of flows per second and Tbps-level throughput. As shown in Figure \ref{fig:scale}, \sys's macro-F1 score experiences only a minor decrease, with about a 13.2\% reduction at the largest scale.
\label{sec:eva_latency}

\vspace{-1em}
\subsection{Latency Microbenchmark.}
\vspace{-0.5em}

\begin{figure}[t]
\vspace{-0.2cm}
    \centering
    \includegraphics[width=\linewidth]{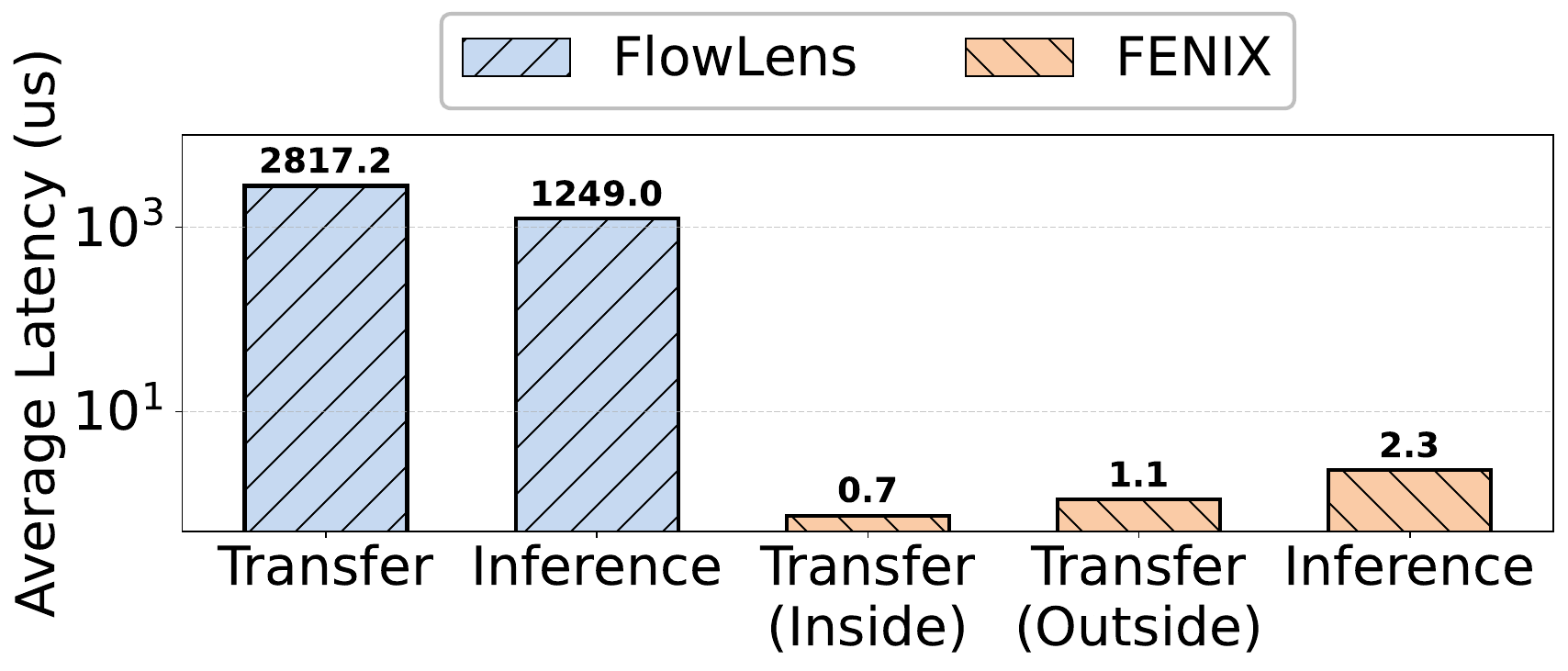}
    \caption{Latency Compare of \sys and FlowLens.}
    \label{fig:latency}
    \vspace{-0.8cm}
\end{figure}

We conduct a comprehensive latency breakdown comparing the control-plane inference approach (FlowLens~\cite{DBLP:conf/ndss/flowlens}) with our proposed \sys, as summarized in Figure~\ref{fig:latency}. To capture microsecond-scale delays, we use an RTT-based method: internal transmission is measured via PCB interconnects, and external transmission via optical modules. Inference latency reflects the total model execution time for each system.

The results demonstrate that \sys significantly outperforms FlowLens across all latency components. FlowLens, relying on general-purpose CPUs and control-plane communication, experiences transmission and inference delays in the millisecond range ($2.1$~ms for transmission and $1.5$~ms for inference). In contrast, \sys benefits from a tightly integrated FPGA architecture where the model and data engines communicate directly within the FPGA fabric, eliminating PCIe and memory bus overhead. This design achieves sub-microsecond internal transmission and $1\text{--}3$$\mu$s external transmission. Most importantly, \sys reduces inference latency by nearly three orders of magnitude, completing inference in just $1.2\mu$s on average compared to FlowLens' $1,000+\mu$s. This improvement stems from in-network FPGA acceleration that offloads model execution from the host CPU and enables efficient, microsecond-scale inference.

\vspace{-1em}
\section{Discussion}
\vspace{-0.5em}
\noindent\textbf{Hardware Adaptability Beyond Tofino.}
\sys is prototyped using a Tofino 2 switch and Xilinx ZU19EG FPGA, but its architecture is fundamentally portable. The discontinuation of Intel's Tofino line~\cite{intel_pcn} does not threaten the broader applicability of \sys, as the programmable switch ecosystem remains rich, including Juniper Trio~\cite{yang2022usingtrio}, Broadcom Trident~\cite{broadcom_bcm78800}, runtime programmable switch designs~\cite{feng2022enabling, xing2022runtime, sigcomm24}, and RMT-based research designs such as dRMT~\cite{chole2017drmt} and Menshen~\cite{wang2022isolation}. 
\added[]{For example, Broadcom Trident 5's NetGNT integrates neural traffic analysis directly into switching silicon. However, publicly available information on such commercial products lacks detailed specifications of AI accelerator capabilities and programming interfaces, making it difficult for the research community to leverage these platforms for academic research and innovation. The hardware and software co-design of \sys anticipates industry trends and provides a platform that can be easily adapted to new generations of programmable switches.}

\noindent\textbf{Prototyping for ASIC Evolution.}
While custom ASICs like Taurus~\cite{swamy_taurus_2022} have set a precedent for per-packet ML inference in data planes—often validated using novel chips like Plasticine~\cite{prabhakar2017plasticine}—barriers to ASIC adoption remain high due to inflexible design and long development cycles. \sys addresses this gap by providing a commercial, reprogrammable prototype that achieves high throughput and low latency at switch nodes, serving as both a practical solution and a reference for future ASIC designers. By demonstrating how DNN inference can be efficiently distributed across programmable switches and FPGAs, \sys charts a pragmatic path toward intelligent data planes in production settings.

\noindent\textbf{Automation and Future Directions.}
Recent frameworks such as Homunculus~\cite{swamy2023homunculus} highlight the growing need for automated ML pipelines and model deployment in data plane environments~\cite{sigcomm23}. While \sys currently requires expert-driven integration when migrating to new tasks or hardware, its modular Model Engine design is well-suited for future automation. However, the current FPGA implementation has inherent limitations regarding flexibility. Layer parameters and feature dimensions are fixed during synthesis, meaning that any DNN architecture updates or changes typically require complete hardware re-synthesis, which limits system adaptability. Our current design primarily supports common CNN and RNN models. Dynamic model updates and partial reconfiguration mechanisms can further enhance system flexibility, and these represent future research directions. Extending \sys with high-level, task-agnostic abstractions, automated toolchains for FPGA and switch targets, and runtime dynamic model switching capabilities would further lower deployment barriers and accelerate its adoption across diverse operational scenarios.
\vspace{-1em}
\section{Related Work}
\vspace{-0.5em}
\label{sec:related}

\noindent\textbf{ML for Network Traffic Analysis.}
Machine learning has become a core tool for network traffic analysis, powering solutions for malicious traffic detection~\cite{kitsune,amado2024peregrinemlbasedmalicioustraffic,fu2021realtime,qing2023low}, website fingerprinting~\cite{DBLP:conf/ndss/RimmerPJGJ18,deng2023robust,shen2020fine}, and fine-grained traffic classification~\cite{al2016adaptive,shen2021accurate}. The field is rapidly advancing toward sophisticated models and tasks, including traffic prediction and encrypted flow analysis~\cite{et-bert,zhao2023yet,zhou2024trafficformer}. However, deploying models at line rate in data plane remains challenging due to hardware constraints, motivating innovations in model compression, quantization, and hardware-aware co-design.

\noindent\textbf{Data Plane Device Roles and Limitations.}
Modern network infrastructures rely on programmable switches, SmartNICs, and FPGAs—each with distinct strengths. SmartNICs efficiently offload compute from servers for functions such as rate limiting, traffic analysis, and protocol acceleration~\cite{radhakrishnan2014senic,siracusano_re-architecting_2022,siracusano_running_2020,arashloo2020enabling,wang2022dip}. Programmable switches provide unparalleled throughput and enable in-network tasks such as DDoS defense and real-time classification~\cite{alcoz2022aggregate,DBLP:conf/uss/jaqen,DBLP:conf/uss/netwarden,li2021enabling,yan2024brain,DBLP:conf/uss/netbeacon,pegasus,lee_switchtree_2020,akem2024jewel,jafri_leo_2024,xie_mousika_2022,busse2019pforest,akem2023flowrest}. FPGAs serve as both prototyping platforms and as flexible data plane accelerators, often complementing the rigid resource budgets of ASIC-based switches~\cite{huang2024spfc, prednsdi25, huang2024diffecn,switchblade, sivaraman2016programmable, LiSIGCOMM2020, swamy_taurus_2022,feng2024f3,scazzariello2023high,zeng2022tiara}.

\noindent\textbf{Comparison and Positioning.}\added[]{
Several commercial products integrate hardware accelerators (FPGAs, DPUs, or AI accelerators) with programmable switch ASICs, such as APS Networks' APS6120Q~\cite{aps6120q} (Intel Tofino 3.2 Tbps ASIC with dual Intel Stratix 10 MX FPGAs) for broadband gateways and network security, and Cisco's smart switch series~\cite{cisco_smart} for enterprise networking. However, these products do not publicly disclose accelerator-switch integration details, deployable model scales, or resource allocation strategies for network intelligence tasks. \sys differs by providing a complete co-design methodology specifically for high-throughput DNN inference in the data plane, addressing the throughput gap challenge between heterogeneous chips and conducting real-world evaluations. Prior work either deploys quantized models in switches (sacrificing accuracy) or executes high-precision models at network edges (introducing latency). \sys leverages tight FPGA-switch integration at central nodes to achieve practical trade-offs between inference accuracy and speed, demonstrating that advanced neural inference can be effectively integrated into switch-centric architectures. This work provides design insights that could inform both commercial deployments and future ASIC developments for intelligent network data planes. We envision these commercial products could serve as potential deployment platforms for \sys, further advancing intelligent network technologies.}
\vspace{-1.2em}
\section{CONCLUSION}
\vspace{-0.7em}
In this paper, we present \sys, a hybrid in-network machine learning system designed to enable fast and accurate traffic analysis directly in the data plane. \sys splits the workload by performing feature extraction on programmable switches and delegating DNN inference to FPGAs, allowing the system to achieve both high throughput and low latency without sacrificing accuracy. To address the bandwidth gap between switches and FPGAs, \sys introduces a token bucket-based mechanism to efficiently regulate feature transmission. Our implementation and evaluation on real-world traffic show that \sys delivers microsecond-level inference latency, multi-terabit throughput, and over 90\% classification accuracy with minimal hardware overhead. Our results serve as a compelling proof-of-concept for the viability of future intelligent data-plane switches.

\vspace{-1em}
\section*{Acknowledgements}
\vspace{-0.7em}
We thank the anonymous reviewers and our shepherd, Eric Keller, for their insightful comments and suggestions. This work is supported by the National Science Foundation for Distinguished Young Scholars of China under Grant No. 62425201; the Science Fund for Creative Research Groups of the National Natural Science Foundation of China under Grant No. 62221003; the National Natural Science Foundation of China under Grant Nos. 62472240, 62394322, U22B2031, 62202473, and 62572473; the Taishan Scholar Foundation of Shandong Province under Grant No. tstp20250724; and the Beijing National Research Center for Information Science and Technology under Grant No. BNR2025RC01010. Ke Xu (xuke@tsinghua.edu.cn) and Su Yao (yaosu@tsinghua.edu.cn) are the corresponding authors.

\clearpage
\bibliographystyle{plain} 
\bibliography{main}

\begin{thebibliography}{10}

\bibitem{amd_zynq_ultrascale_plus_mpsoc}
{Advanced Micro Devices (AMD)}.
\newblock {AMD Zynq™ UltraScale+™ MPSoC}.
\newblock \url{https://www.amd.com/en/products/adaptive-socs-and-fpgas/soc/zynq-ultrascale-plus-mpsoc.html}.
\newblock Accessed: 2025-01.

\bibitem{akem2024jewel}
Aristide Tanyi-Jong Akem, Beyza B{\"u}t{\"u}n, Michele Gucciardo, Marco Fiore, et~al.
\newblock Jewel: Resource-efficient joint packet and flow level inference programmable switches.
\newblock In {\em IEEE International Conference on Computer Communications (INFOCOM)}, 2024.

\bibitem{akem2023flowrest}
Aristide Tanyi-Jong Akem, Michele Gucciardo, and Marco Fiore.
\newblock Flowrest: Practical flow-level inference programmable switches with random forests.
\newblock In {\em IEEE International Conference on Computer Communications (INFOCOM)}, pages 1--10. IEEE, 2023.

\bibitem{al2016adaptive}
Khaled Al-Naami, Swarup Chandra, Ahmad Mustafa, Latifur Khan, Zhiqiang Lin, and Bhavani Hamlen, Kevand~Thuraisingham.
\newblock Adaptive encrypted traffic fingerprinting with bi-directional dependence.
\newblock In {\em Proceedings of the 32nd Annual Conference on Computer Security Applications}, pages 177--188, 2016.

\bibitem{alcoz2022aggregate}
Albert~Gran Alcoz, Vincent Strohmeier, Martand~Lenders, and Laurent Vanbever.
\newblock Aggregate-based congestion control for pulse-wave ddos defense.
\newblock In {\em Proceedings of the Annual Conference of the ACM Special Interest Group on Data Communication (SIGCOMM)}, pages 693--706, 2022.

\bibitem{amado2024peregrinemlbasedmalicioustraffic}
João~Romeiras Amado, Francisco Pereira, David Pissarra, Salvatore Signorello, Miguel Correia, and Fernando M.~V. Ramos.
\newblock Peregrine: Ml-based malicious traffic detection for terabit networks.
\newblock {\em arXiv preprint arXiv:2403.18788}, 2024.

\bibitem{AMD2023VitisAI}
{AMD}.
\newblock {\em Vitis AI 3.5 Documentation}.
\newblock Accessed: 2025-01.

\bibitem{AMD2024VitisHLS}
{AMD}.
\newblock {\em Vitis High-Level Synthesis User Guide (UG1399)}.
\newblock Accessed: 2025-01.

\bibitem{switchblade}
Muhammad~Bilal Anwer, Murtaza Motiwala, Muhammad~Mukarram BTariq, and Nick Feamster.
\newblock Switchblade: a platform for rapid deployment of network protocols.
\newblock In {\em Proceedings of the Annual Conference of the ACM Special Interest Group on Data Communication (SIGCOMM)}, pages 183--194, 2010.

\bibitem{aps6120q}
{APS Networks GmbH}.
\newblock {APS6120Q Advanced Programmable Switch}.
\newblock \url{https://www.aps-networks.com/wp-content/uploads/2021/06/210616_APS6120Q_preliminary.pdf}, 2021.
\newblock Product Datasheet, Accessed: 2025-10.

\bibitem{arashloo2020enabling}
Mina~Tahmasbi Arashloo, Alexey Lavrov, Manya Ghobadi, Jennifer Rexford, David Walker, and David Wentzlaff.
\newblock Enabling programmable transport protocols high-speed nics.
\newblock In {\em USENIX Symposium on Networked Systems Design and Implementation (NSDI)}, pages 93--109, 2020.

\bibitem{DBLP:conf/ndss/flowlens}
Diogo Barradas, Nuno Santos, Lu{\'{\i}}s Rodrigues, Salvatore Signorello, Fernando M.~V. Ramos, and Andr{\'{e}} Madeira.
\newblock Flowlens: Enabling efficient flow classification for ml-based network security applications.
\newblock In {\em Network and Distributed System Security Symposium (NDSS)}, 2021.

\bibitem{bosshart_p4_2014}
Pat Bosshart, Dan Daly, Glen Gibb, Nick Izzard, Martand~McKeown, Jennifer Rexford, Cole Schlesinger, Dan Talayco, George Vahdat, Amand~Varghese, and {others}.
\newblock P4: {Programming} protocol-independent packet processors.
\newblock {\em ACM SIGCOMM Computer Communication Review}, 44(3):87--95, 2014.

\bibitem{busse2019pforest}
Coralie Busse-Grawitz, Roland Meier, Alexander Dietm{\"u}ller, Tobias B{\"u}hler, and Laurent Vanbever.
\newblock pforest: In-network inference with random forests.
\newblock {\em arXiv preprint arXiv:1909.05680}, 2019.

\bibitem{chole2017drmt}
Sharad Chole, Andy Fingerhut, Sha Ma, Anirudh Sivaraman, Shay Vargaftik, Alon Berger, Gal Mendelson, Mohammad Alizadeh, Shang-Tse Chuang, Isaac Keslassy, et~al.
\newblock drmt: Disaggregated programmable switching.
\newblock In {\em Proceedings of the Annual Conference of the ACM Special Interest Group on Data Communication (SIGCOMM)}, pages 1--14, 2017.

\bibitem{cisco_smart}
{Cisco Systems}.
\newblock {Cisco N9300 Series Smart Switches}.
\newblock \url{https://www.cisco.com/site/us/en/products/networking/cloud-networking-switches/9300-series-smart-switches/index.html}.
\newblock Accessed: 2025-10.

\bibitem{deng2023robust}
Xinhao Deng, Qilei Yin, Zhuotao Liu, Xiyuan Zhao, Qi~Li, Mingwei Xu, Ke~Xu, and Jianping Wu.
\newblock Robust multi-tab website fingerprinting attacks the wild.
\newblock In {\em IEEE Symposium on Security and Privacy (S\&P)}, pages 1005--1022, 2023.

\bibitem{dpdk_2025}
{DPDK Project}.
\newblock {DPDK - The open source data plane development kit accelerating network performance}.
\newblock \url{https://www.dpdk.org/}.
\newblock Accessed: 2025-01.

\bibitem{prednsdi25}
Xinle Du, Tong Li, Guangmeng Zhou, Zhuotao Liu, Hanlin Huang, Xiangyu Gao, Mowei Wang, kun Tan, and ke~Xu.
\newblock Pred: Performance-oriented random early detection for consistently stable performance in datacenters.
\newblock In {\em USENIX Symposium on Networked Systems Design and Implementation (NSDI)}, pages 1--20, 2025.

\bibitem{feng2024f3}
Weiqi Feng, Jiaqi Gao, Xiaoqi Chen, Gianni Antichi, Ran~Ben Basat, Michael~Mingchao Shao, Ying Zhang, and Minlan Yu.
\newblock F3: Fast and flexible network telemetry with an fpga coprocessor.
\newblock {\em Proceedings of the ACM on Networking}, 2(CoNEXT4):1--22, 2024.

\bibitem{feng2022enabling}
Yong Feng, Zhikang Chen, Haoyu Song, Wenquan Xu, Jiahao Li, Zijian Zhang, Tong Yun, Ying Wan, and Bin Liu.
\newblock Enabling in-situ programmability in network data plane: From architecture to language.
\newblock In {\em USENIX Symposium on Networked Systems Design and Implementation (NSDI)}, pages 635--649, 2022.

\bibitem{sigcomm24}
Yong Feng, Zhikang Chen, Haoyu Song, Yinchao Zhang, Hanyi Zhou, Ruoyu Sun, Wenkuo Dong, Peng Lu, Shuxin Liu, Chuwen Zhang, et~al.
\newblock Empower programmable pipeline for advanced stateful packet processing.
\newblock In {\em USENIX Symposium on Networked Systems Design and Implementation (NSDI)}, pages 491--508, 2024.

\bibitem{firestone2018azure}
Daniel Firestone, Andrew Putnam, Sambhrama Mundkur, Derek Chiou, Alireza Dabagh, Mike Andrewartha, Hari Angepat, Vivek Bhanu, Adrian Caulfield, Eric Chung, et~al.
\newblock Azure accelerated networking:smartnics the public cloud.
\newblock In {\em USENIX Symposium on Networked Systems Design and Implementation (NSDI)}, pages 51--66, 2018.

\bibitem{fu2021realtime}
Chuanpu Fu, Qi~Li, Meng Shen, and Ke~Xu.
\newblock Realtime robust malicious traffic detection via frequency domaanalysis.
\newblock In {\em Proceedings of the ACM SIGSAC Conference on Computer and Communications Security (CCS)}, pages 3431--3446, 2021.

\bibitem{iscxvpn}
Gerard~Drapper Gil, Arash~Habibi Lashkari, Mohammad Mamun, and Ali~A Ghorbani.
\newblock Characterization of encrypted and vpn traffic using time-related features.
\newblock In {\em Proceedings of the 2nd international conference on information systems security and privacy (ICISSP)}, pages 407--414, 2016.

\bibitem{gupta_loom_2019}
Shreyan Gupta, Jiacheng He, Vamsi Olagappan, Arvind Raghunathan, Ausav Foong, Konstantina Papagiannaki, and Jitendra Padhye.
\newblock Loom: Flexible and efficient {NIC} packet scheduling.
\newblock In {\em USENIX Symposium on Networked Systems Design and Implementation (NSDI)}, pages 73--88, 2019.

\bibitem{huang2024spfc}
Hanlin Huang, Xinle Du, Tong Li, Haiyang Wang, Ke~Xu, Mowei Wang, and Huichen Dai.
\newblock Re-architecting buffer management in lossless ethernet.
\newblock {\em IEEE/ACM Transactions on Networking}, 32(6):4749--4764, 2024.

\bibitem{huang2024diffecn}
Hanlin Huang, Ke~Xu, Tong Li, Zhuotao Liu, Xinle Du, and Xiangyu Gao.
\newblock Diffecn: Differential ecn marking for datacenter networks.
\newblock {\em IEEE/ACM Transactions on Networking}, 33(1):210--225, 2025.

\bibitem{broadcom_bcm78800}
Broadcom Inc.
\newblock Bcm78800: Strataxgs trident 5 ethernet switch series.
\newblock \url{https://www.broadcom.com/products/ethernet-connectivity/switching/strataxgs/bcm78800}.
\newblock Accessed: 2025-01.

\bibitem{tofino}
Intel.
\newblock Intel® intelligent fabric processors.
\newblock \url{https://www.intel.com/content/www/us/en/products/details/network-io/intelligent-fabric-processors.html}.
\newblock Accessed: 2025-01.

\bibitem{tofino2}
{Intel}.
\newblock Intel® tofino™ 2.
\newblock \url{https://www.intel.com/content/www/us/en/products/sku/218648/intel-tofino-2-12-8-tbps-20-stage-4-pipelines/specifications.html}.
\newblock Accessed: 2025-01.

\bibitem{intel_pcn}
Intel.
\newblock Product change notification: Tofino 2.
\newblock \url{https://cdrdv2-public.intel.com/827577/PCN827577-00.pdf}.
\newblock Accessed: 2025-01.

\bibitem{jafri_leo_2024}
Syed~Usman Jafri, Sanjay Rao, Vishal Shrivastav, and Mohit Tawarmalani.
\newblock Leo: {Online} {ML}-based {Traffic} {Classification} at {Multi}-{Terabit} {Line} {Rate}.
\newblock In {\em USENIX Symposium on Networked Systems Design and Implementation (NSDI)}, pages 1573--1591, 2024.

\bibitem{kim2025sketchfeature}
Sian Kim, Seyed Mohammad~Mehdi Mirnajafizadeh, Bara Kim, Rhongho Jang, and D~Nyang.
\newblock Sketchfeature: High-quality per-flow feature extractor towards security-aware data plane.
\newblock In {\em Network and Distributed System Security Symposium (NDSS)}, 2025.

\bibitem{lee_switchtree_2020}
Jong-Hyouk Lee and Kamal Singh.
\newblock Switchtree: in-network computing and traffic analyses with random forests.
\newblock {\em Neural Computing and Applications}, pages 1--12, 2020.

\bibitem{li2021enabling}
Guanyu Li, Menghao Zhang, Shicheng Wang, Chang Liu, Mingwei Xu, Ang Chen, Guofei Hu, Hongxand~Gu, Qi~Li, and Jianping Wu.
\newblock Enabling performant, flexible and cost-efficient ddos defense with programmable switches.
\newblock {\em IEEE/ACM Transactions on Networking}, 29(4):1509--1526, 2021.

\bibitem{LiSIGCOMM2020}
Tong Li, Kai Zheng, Ke~Xu, Rahul~Arvind Jadhav, Tao Xiong, Keith Winstein, and Kun Tan.
\newblock {TACK:} improving wireless transport performance by taming acknowledgments.
\newblock In {\em Proceedings of the Annual Conference of the ACM Special Interest Group on Data Communication (SIGCOMM)}, pages 15--30, 2020.

\bibitem{et-bert}
Xinjie Lin, Gang Xiong, Gaopeng Gou, Zhen Li, Junzheng Shi, and Jing Yu.
\newblock Et-bert: A contextualized datagram representation with pre-training transformers for encrypted traffic classification.
\newblock In {\em Proceedings of the ACM Web Conference (WWW)}, 2022.

\bibitem{DBLP:conf/uss/jaqen}
Zaoxing Liu, Hun Namkung, Georgios Nikolaidis, Jeongkeun Lee, Changhoon Kim, XJand~Vladimir Braverman, Minlan Yu, and Vyas Sekar.
\newblock Jaqen: {A} high-performance switch-native approach for detecting and mitigating volumetric ddos attacks with programmable switches.
\newblock In {\em {USENIX} Security Symposium (USENIX Security)}, pages 3829--3846, 2021.

\bibitem{kitsune}
Yisroel Mirsky, Tomer Doitshman, Yuval Elovici, and Asaf Shabtai.
\newblock Kitsune: An ensemble of autoencoders for online network intrusion detection.
\newblock In {\em Network and Distributed System Security Symposium (NDSS)}, 2018.

\bibitem{noauthor_netfpga_nodate}
{NetFPGA}.
\newblock {NetFPGA}.
\newblock \url{https://netfpga.org/}.
\newblock Accessed: 2025-01.

\bibitem{nvidia_connectx7_2025}
{NVIDIA}.
\newblock {NVIDIA ConnectX-7 SmartNICs}.
\newblock \url{https://www.nvidia.cn/networking/ethernet-adapters/}.
\newblock Accessed: 2025-01.

\bibitem{prabhakar2017plasticine}
Raghu Prabhakar, Yaqi Zhang, David Koeplinger, Matt Feldman, Tian Zhao, Stefan Hadjis, Ardavan Pedram, Christos Kozyrakis, and Kunle Olukotun.
\newblock Plasticine: A reconfigurable architecture for parallel paterns.
\newblock {\em ACM SIGARCH Computer Architecture News}, 45(2):389--402, 2017.

\bibitem{qing2023low}
Yuqi Qing, Qilei Yin, Xinhao Deng, Yihao Chen, Zhuotao Liu, Kun Sun, Ke~Xu, Jia Zhang, and Qi~Li.
\newblock Low-quality training data only? a robust framework for detecting encrypted malicious network traffic.
\newblock In {\em Network and Distributed System Security Symposium (NDSS)}, 2024.

\bibitem{radhakrishnan2014senic}
Sivasankar Radhakrishnan, Yilong Geng, Vimalkumar Jeyakumar, Abdul Kabbani, George Porter, and Amin Vahdat.
\newblock Senic: Scalable nic for end-host rate limiting.
\newblock In {\em USENIX Symposium on Networked Systems Design and Implementation (NSDI)}, pages 475--488, 2014.

\bibitem{DBLP:conf/ndss/RimmerPJGJ18}
Vera Rimmer, Davy Preuveneers, Marc Juarez, Tom van Goethem, and Wouter Joosen.
\newblock Automated website fingerprinting through deep learning.
\newblock In {\em Network and Distributed System Security Symposium (NDSS)}, 2018.

\bibitem{scazzariello2023high}
Mariano Scazzariello, Tommaso Caiazzi, Hamid Ghasemirahni, Tom Barbette, Dejan Kosti{\'c}, and Marco Chiesa.
\newblock A high-speed stateful packet processing approach for tbps programmable switches.
\newblock In {\em USENIX Symposium on Networked Systems Design and Implementation (NSDI)}, pages 1237--1255, 2023.

\bibitem{shen2020fine}
Meng Shen, Yiting Liu, Liehuang Zhu, Xiaojiang Du, and Jiankun Hu.
\newblock Fine-grained webpage fingerprinting using only packet length information of encrypted traffic.
\newblock {\em IEEE Transactions on Information Forensics and Security (TIFS)}, 16:2046--2059, 2020.

\bibitem{shen2021accurate}
Meng Shen, Jinpeng Zhang, Liehuang Zhu, Ke~Xu, and Xiaojiang Du.
\newblock Accurate decentralized application identification via encrypted traffic analysis using graph neural networks.
\newblock {\em IEEE Transactions on Information Forensics and Security (TIFS)}, 16:2367--2380, 2021.

\bibitem{siracusano_re-architecting_2022}
Giuseppe Siracusano, Salvator Galea, Davide Sanvito, Mohammad Malekzadeh, Gianni Antichi, Paolo Costa, Hamed Haddadi, and Roberto Bifulco.
\newblock Re-architecting traffic analysis with neural network interface cards.
\newblock In {\em USENIX Symposium on Networked Systems Design and Implementation (NSDI)}, pages 513--533, 2022.

\bibitem{siracusano_running_2020}
Giuseppe Siracusano, Salvator Galea, Davide Sanvito, Mohammad Malekzadeh, Hamed Haddadi, Gianni Antichi, and Roberto Bifulco.
\newblock Running neural networks on the {NIC}.
\newblock {\em arXiv preprint arXiv:2009.02353}, 2020.

\bibitem{sivaraman2016programmable}
Anirudh Sivaraman, Suvinay Subramanian, Mohammad Alizadeh, Sharad Chole, Shang-Tse Chuang, Anurag Agrawal, Hari Balakrishnan, Tom Edsall, and Nick Katti, Sachand~McKeown.
\newblock Programmable packet scheduling at line rate.
\newblock In {\em Proceedings of the Annual Conference of the ACM Special Interest Group on Data Communication (SIGCOMM)}, pages 44--57, 2016.

\bibitem{spang2019estimating}
Bruce Spang and Nick McKeown.
\newblock On estimating the number of flows.
\newblock In {\em Stanford Workshop on Buffer Sizing}, 2019.

\bibitem{swamy_taurus_2022}
Tushar Swamy, Alexander Rucker, Muhammad Shahbaz, Ishan Gaur, and Kunle Olukotun.
\newblock Taurus: a data plane architecture for per-packet {ML}.
\newblock In {\em Proceedings of the 27th {ACM} {International} {Conference} on {Architectural} {Support} for {Programming} {Languages} and {Operating} {Systems} (ASPLOS)}, 2022.

\bibitem{swamy2023homunculus}
Tushar Swamy, Annus Zulfiqar, Luigi Nardi, Muhammad Shahbaz, and Kunle Olukotun.
\newblock Homunculus: Auto-generating efficient data-plane ml pipelines for datacenter networks.
\newblock In {\em Proceedings of the 28th {ACM} {International} {Conference} on {Architectural} {Support} for {Programming} {Languages} and {Operating} {Systems} (ASPLOS)}, pages 329--342, 2023.

\bibitem{wang2022isolation}
Tao Wang, Xiangrui Yang, Gianni Antichi, Anirudh Sivaraman, and Aurojit Panda.
\newblock Isolation mechanisms for high-speedpacket-processing pipelines.
\newblock In {\em USENIX Symposium on Networked Systems Design and Implementation (NSDI)}, pages 1289--1305, 2022.

\bibitem{ustc}
Wei Wang, Ming Zhu, Xuewen Zeng, Xiaozhou Ye, and Yiqiang Sheng.
\newblock Malware traffic classification using convolutional neural network for representation learning.
\newblock In {\em International conference on information networking (ICOIN)}, pages 712--717, 2017.

\bibitem{wang2022dip}
Ziqiang Wang, Zhuotao Liu, Xiaoliang Wang, Songtao Fu, and Ke~Xu.
\newblock Dip: unifying network layer innovations using shared l3 core functions.
\newblock In {\em Proceedings of the 21st ACM Workshop on Hot Topics Networks (HotNets)}, pages 60--67, 2022.

\bibitem{xgboost_documentation_2025}
{XGBoost Developers}.
\newblock {Introduction to XGBoost Python Package}.
\newblock \url{https://xgboost.readthedocs.io/en/stable/python/python_intro.html}.
\newblock Accessed: 2025-01.

\bibitem{xie_mousika_2022}
Guorui Xie, Qing Li, Yutao Dong, Yong Duan, Guangland~Jiang, and Jingpu Duan.
\newblock Mousika: {Enable} {General} {In}-{Network} {Intelligence} {Programmable} {Switches} by {Knowledge} {Distillation}.
\newblock In {\em IEEE International Conference on Computer Communications (INFOCOM)}, pages 1938--1947, 2022.

\bibitem{xing2022runtime}
Jiarong Xing, Kuo-Feng Hsu, Matty Kadosh, Alan Lo, Yonatan Piasetzky, Arvind Krishnamurthy, and Ang Chen.
\newblock Runtime programmable switches.
\newblock In {\em USENIX Symposium on Networked Systems Design and Implementation (NSDI)}, pages 651--665, 2022.

\bibitem{DBLP:conf/uss/netwarden}
Jiarong Xing, Qiao Kang, and Ang Chen.
\newblock Netwarden: Mitigating network covert channels while preserving performance.
\newblock In {\em {USENIX} Security Symposium (USENIX Security)}, pages 2039--2056, 2020.

\bibitem{sigcomm23}
Wenquan Xu, Zijian Zhang, Yong Feng, Haoyu Song, Zhikang Chen, Wenfei Wu, Guyue Liu, Yinchao Zhang, Shuxin Liu, Zerui Tian, and Bin Liu.
\newblock Clickinc: In-network computing as a service in heterogeneous programmable data-center networks.
\newblock In {\em Proceedings of the Annual Conference of the ACM Special Interest Group on Data Communication (SIGCOMM)}, page 798–815, 2023.

\bibitem{yan2024brain}
Jinzhu Yan, Haotian Xu, Zhuotao Liu, Qi~Li, Ke~Xu, Mingwei Xu, and Jianping Wu.
\newblock Brain-on-switch: Towards advanced intelligent network data plane via nn-driven traffic analysis at line-speed.
\newblock In {\em USENIX Symposium on Networked Systems Design and Implementation (NSDI)}, pages 419--440, 2024.

\bibitem{yang2022usingtrio}
Mingran Yang, Alex Baban, Valery Kugel, Jeff Libby, Scott Mackie, Swamy Sadashivaiah~Renu Kananda, Chang-Hong Wu, and Manya Ghobadi.
\newblock Using trio: juniper networks' programmable chipset-for emerging in-network applications.
\newblock In {\em Proceedings of the Annual Conference of the ACM Special Interest Group on Data Communication (SIGCOMM)}, pages 633--648, 2022.

\bibitem{zeng2022tiara}
Chaoliang Zeng, Layong Luo, Teng Zhang, Zilong Wang, Luyang Li, Wenchen Han, Nan Chen, Lebing Wan, Lichao Liu, Zhipeng Ding, et~al.
\newblock Tiara: A scalable and efficient hardware acceleration architecture for stateful layer-4 load balancing.
\newblock In {\em USENIX Symposium on Networked Systems Design and Implementation (NSDI)}, pages 1345--1358, 2022.

\bibitem{pegasus}
Yinchao Zhang, Su~Yao, Yong Feng, Kang Chen, Tong Li, Zhuotao Liu, Yi~Zhao, Lexuan Zhang, Xiangyu Gao, Feng Xiong, Qi~Li, and Ke~Xu.
\newblock Pegasus: A universal framework for scalable deep learning inference on the dataplane.
\newblock In {\em Proceedings of the Annual Conference of the ACM Special Interest Group on Data Communication (SIGCOMM)}, page 692–706, 2025.

\bibitem{zhao2023yet}
Ruijie Zhao, Mingwei Zhan, Xianwen Deng, Yanhao Wang, Yijun Wang, Guan Gui, and Zhi Xue.
\newblock Yet another traffic classifier: A masked autoencoder based traffic transformer with multi-level flow representation.
\newblock In {\em Proceedings of the AAAI Conference on Artificial Intelligence}, pages 5420--5427, 2023.

\bibitem{zheng_iisy_2022}
Changgang Zheng, Zhaoqi Xiong, Thanh~T Bui, Siim Kaupmees, Riyad Bensoussane, Antoine Bernabeu, Shay Vargaftik, Yaniv Ben-Itzhak, and Noa Zilberman.
\newblock {IIsy}: {Practical} in-network classification.
\newblock {\em arXiv preprint arXiv:2205.08243}, 2022.

\bibitem{zhou2024trafficformer}
Guangmeng Zhou, Xiongwen Guo, Zhuotao Liu, Tong Li, Qi~Li, and Ke~Xu.
\newblock Trafficformer: An efficient pre-trained model for traffic data.
\newblock In {\em 2025 IEEE Symposium on Security and Privacy (S\&P)}, pages 102--102, 2024.

\bibitem{DBLP:conf/uss/netbeacon}
Guangmeng Zhou, Zhuotao Liu, Chuanpu Fu, Qi~Li, and Ke~Xu.
\newblock An efficient design of intelligent network data plane.
\newblock In {\em {USENIX} Security Symposium (USENIX Security)}, pages 6203--6220, 2023.

\end{thebibliography}
\clearpage
\appendix

\end{document}